\documentclass[11pt]{article}
\usepackage{mymacros}

\title{Genuine Multi-Entropy of Fully Symmetric Gaussian States}

\author[a,b]{Hugo A. Camargo,}
\affiliation[a]{Physics Division, National Center for Theoretical Sciences, National Taiwan University, Taipei 106319, Taiwan}
\affiliation[b]{Department of Physics, National Sun Yat-sen University, Kaohsiung 80424, Taiwan}
\author[c]{and Mitsuhiro Nishida}
\affiliation[c]{National Institute of Technology, Yuge College, Ehime 794-2593, Japan}
\emailAdd{  hugo.camargo@phys.ncts.ntu.edu.tw, mnishida124@gmail.com }

\date{October 2026}


\begin{document}

\abstract{We study the tri-partite genuine Rényi multi-entropy of fully symmetric Gaussian states as a function of the squeezing parameter for different total number of bosonic modes under different subsystem configurations. Exact results for a few modes are obtained, and we conjecture an asymptotic formula at the limit of large squeezing parameter as a function of the replica index $n$, while also finding subleading subsystem-size dependent terms for $n=3$ under a specific tripartition. We analytically show that the genuine $n=2$ Rényi multi-entropy of any pure bosonic Gaussian states is zero for any number of modes and for any tripartition. Furthermore, to investigate the dependence on subsystem size, we plot the curve of genuine Rényi multi-entropy as a function of the number of modes in the subsystems while keeping the total number of modes fixed. We discuss differences and similarities between this result and studies of the multi-entropy curve for Haar random states. In particular, we identify a subsystem size configuration related with the multi-entropy time in an evaporating black hole for which the behavior of the genuine multi-entropy changes.}

\maketitle

\section{Introduction}
\label{sec:Intro}

The characterization of entanglement has led to vast developments in the study of quantum many-body and topological systems~\cite{Vidal:2002rm,Kitaev:2005dm,Eisert:2008ur,Horodecki:2009zz}, conformal and quantum field theories~\cite{Holzhey:1994we,Calabrese:2004eu,Calabrese:2009qy}, and in holography~\cite{Ryu:2006bv,Ryu:2006ef,Hubeny:2007xt,Casini:2011kv,Nishioka:2018khk}. In particular, in the latter context, holographic entanglement entropy has emerged as a powerful tool for diagnosing the emergence of bulk semiclassical spacetime from the non-local physics of boundary subregions in strongly-coupled quantum systems. This, in turn, has led to a deeper understanding of black hole entropy~\cite{Bekenstein:1973ur,Srednicki:1993im} and the black hole information paradox~\cite{Hawking:1975vcx}, which are some of the biggest mysteries in our fundamental understanding of quantum gravity~\footnote{See e.g.~\cite{Almheiri:2020cfm} for a review of recent developments in the black hole information paradox.}.

These developments rely on bi-partite entanglement measures, such as entanglement entropy~\cite{Bombelli:1986rw} and its Rényi generalizations. An equally important endeavor is to characterize \emph{multi-partite} correlations and the entanglement content of $\mathtt{q}-$party quantum systems~\cite{Walter:2016lgl,Ma:2023ecg,Horodecki:2024bgc}. In general, it is believed that understanding multi-partite entanglement could directly impact quantum computing technologies through measurement-based protocols~\cite{Briegel:2009inn,Wong:2022mnv}. At the same time, there exists increasing evidence that multi-partite entanglement also plays an important role in holography beyond bi-partite entanglement such as in the holographic Markov gap \cite{Akers:2019gcv, Hayden:2021gno}.

However, developing a coherent description of entanglement in multi-partite states pre-sents a greater challenge due to the inherent ambiguity of what ``entanglement'' means in $\mathtt{q}-$party quantum systems. This is because unlike in bi-partite quantum systems, there are different and inequivalent ways of characterizing multi-partite entanglement in $\mathtt{q}-$party systems due to the presence of partially-separable states. Indeed, the first studies of entanglement in multi-partite states highlighted such intricacies, initiating the study of multi-party entanglement~\cite{Svetlichny:1987prd,Greenberger:1989tfe,Dur:2000zz}. 

Recently, a new family of quantities used to diagnose multi-partite entanglement was introduced in~\cite{Gadde:2022cqi,Penington:2022dhr}. These quantities, collectively referred to as \emph{multi-entropy}, are $\mathtt{q}-$party generalizations of the usual bi-partite entanglement entropy that are symmetric with respect to all the parties. To be precise, the $n-$th Rényi $\mathtt{q}-$party multi-entropy, $S^{(\mathtt{q})}_{n}$, defined from $n^{\mathtt{q}-1}$ copies of the density matrix of a $\mathtt{q}-$party system, can be computed through a generalization of the replica method used to compute the $n-$th Rényi entanglement entropy $S^{(\mathtt{2})}_{n}$. Multi-entropy can be seen as a particular example of a larger class of \emph{multi-invariants} corresponding to specific local unitary invariants that can be used to define partial traces with respect to a specific finite group symmetry~\cite{Gadde:2024taa}. Moreover, multi-entropy has been proposed to have concrete geometric realizations~\cite{Gadde:2022cqi,Gadde:2023zzj} in terms of areas of codimension $2$ multi-way cuts in holographic spacetime. Recent efforts have classified the properties of multi-entropy and other multi-invariant measures in a variety of settings; from spin systems \cite{Iizuka:2026qqg,Zhang:2026pkk}, topological phases \cite{Sheffer:2025jtc,Sheffer:2025zyr,Sheffer:2026dgj,Sohal:2026tpv}, quantum field theories (QFTs)~\cite{Harper:2024ker,Harper:2025uui,Yuan:2025dgx,Chang:2026uds}, topological quantum field theories (TQFTs)~\cite{DelZotto:2026fpw,Iizuka:2026mlw}, graph states \cite{Iizuka:2025pqq,Akella:2026xza}, quantum error correction codes~\cite{Akella:2025owv,Akella:2026rbe}, random tensor networks~\cite{Akella:2026bci,Hu:2026bhg}, holography \cite{Yuan:2024yfg,Iizuka:2025elr,Balasubramanian:2025hxg,Ju:2025eyn,Anegawa:2025prn,Chen:2026xtx,Iizuka:2026ahd,Balasubramanian:2026chr,Fujiki:2026qdt,Naskar:2026zka,Chen:2026hcf,Ju:2026zbu,Ju:2026lst,Anegawa:2026yee}, and Haar random states in the context of black hole evaporation~\cite{Iizuka:2024pzm,Iizuka:2025ioc,Iizuka:2025caq}.

One issue with multi-entropy is the analytic continuation to $n=1$. Unlike entanglement entropy, multi-entropy at $n=1$ cannot be defined directly without the analytic continuation. To evaluate it, we need to know the analytic expressions of $S^{(\mathtt{q})}_{n}$ as a function of the replica index $n$, and not simply its numerical values. Examples where the analytic expression can be obtained for any $n$ are the GHZ state \cite{Gadde:2022cqi} and the Lifshitz ground state \cite{Berthiere:2025toi}. It is important to find and study other examples where the analytic expressions of $S^{(\mathtt{q})}_{n}$ can be examined.

The purpose of this manuscript is two-fold: Firstly, motivated by the recent work studying multi-entropy for ground states in free $(1+1)-$dimensional CFTs~\cite{Harper:2025uui}, we perform analytic and numerical computations of multi-entropy for a class of continuous-variable (CV) bosonic Gaussian states that are invariant under the exchange of any two modes and which are completely characterized by the single-mode purity $\mu$, called fully symmetric Gaussian states~\cite{Adesso:2004fz,Adesso:2008lis}. Gaussian states naturally appear as ground states of free quantum field theories (QFTs), while fully symmetric Gaussian states are an important class of CV systems that provide key resources for multi-party CV quantum information protocols~\cite{Braunstein:2005zz,Adesso:2007tx} and that can be experimentally prepared~\cite{vanLoock:1999gba,Adesso:2005asb}. Restricting ourselves to fully symmetric Gaussian states allows us to find closed-form expressions and establish general results in different limits.   

Secondly, we consider the Gaussian-state counterpart of the holographic \emph{multi-entropy curve} describing black hole evaporation studied in~\cite{Iizuka:2024pzm,Iizuka:2025ioc,Iizuka:2025caq}. A version of the Page curve for ensembles of pure fermionic Gaussian states was studied in~\cite{Bianchi:2021lnp}, where it was found that the average entanglement entropy as a function of subsystem size follows a Page-like curve~\cite{Page:1993df}. A natural question is whether it is possible to find a multi-entropy-like curve for fully symmetric Gaussian states as a function of subsystem size and whether it behaves similarly to a multi-entropy-like curve of Haar random states for black hole evaporation.

This manuscript is structured as follows: In Sec.~\ref{sec:Review} we review the construction of \emph{genuine} Rényi multi-entropy GM$_{n}^{(\mathtt{3})}(A:B:C)$ in tri-partite subsystems $\mathcal{H}_{A}\otimes \mathcal{H}_{B}\otimes \mathcal{H}_{C}
$ and discuss its general properties for triangle states $\ket{\Delta}_{ABC}$ and the $d$-dimensional GHZ state $\vert \textrm{GHZ} \rangle$, establishing general expectations for these classes of states. Here, genuine Rényi multi-entropy is a linear combination of Rényi multi-entropy that vanishes for separable states in multi-partite systems, where a nonzero value of genuine Rényi multi-entropy indicates the existence of genuinely multi-partite entanglement. We then discuss the general formalism for describing pure bosonic Gaussian states of $N$ modes and describe the strategy for computing GM$_{n}^{(\mathtt{3})}(A:B:C)$ in this general class of states. In Sec.~\ref{sec:MEGaussian}, we introduce fully symmetric Gaussian states and discuss the behavior of GM$_{n}^{(\mathtt{3})}(A:B:C)$ for different values of $n$ and for different subsystem sizes $N_{A},N_{B},N_{C}$, finding analytic results in different cases and specific limits. In Sec.~\ref{sec:ProogGMn2} we show the vanishing of the $n=2$ genuine Rényi multi-entropy GM$_{2}^{(\mathtt{3})}$ for general bosonic Gaussian states due to a specific factorization property of the $n=2$ and $\mathtt{q}=3$ replica partition function $Z^{(\mathtt{3})}_{2}$. Afterwards, in Sec.~\ref{sec:GaussianMECurve} we discuss the Gaussian state analogue of the multi-entropy curve for GM$_{3}^{(\mathtt{3})}(A:B:C)$ for different values of the squeezing parameter $a=\mu^{-1}$. Finally, in Sec.~\ref{sec:Discussion} we offer our conclusion, discuss our results, and provide some possible future directions. In Appendix~\ref{app1} we discuss more details of the matrix structure for mixed Gaussian states used for the proof in Sec.~\ref{sec:ProogGMn2}, while in Appendix~\ref{app2} we provide additional details for permutation-symmetric quadratic Hamiltonians. 

\section{Review of genuine multi-entropy in tri-partite systems}
\label{sec:Review}

In this section, we review how to construct genuine multi-entropy in tri-partite subsystems. As a simple example, we review the genuine multi-entropy of the GHZ state. We also explain how to formulate and calculate the genuine Rényi multi-entropy of pure bosonic Gaussian states from density matrices. 

\subsection{Genuine Rényi multi-entropy in tri-partite systems}
First, we define Rényi multi-entropy as a generalization of Rényi entropy for multi-partite subsystems. The $n$-th Rényi multi-entropy of a pure state $\ket{\psi}$ on  $\mathtt{q}$-partite subsystems $A_1$, $A_2$, $\dots$, $A_\mathtt{q}$ is defined by \cite{Gadde:2022cqi, Penington:2022dhr, Gadde:2023zzj, Gadde:2023zni}
\begin{align}
\label{thedefinition}
S^{(\mathtt{q})}_n(A_1:A_2:\dots:A_\mathtt{q}) &:= \frac{1}{1-n}\frac{1}{n^{\mathtt{q}-2}}\log \frac{Z^{(\mathtt{q})}_n}{(Z^{(\mathtt{q})}_1)^{n^{\mathtt{q}-1}}},\\
Z^{(\mathtt{q})}_n &:= \bra{\psi}^{\otimes n^{\mathtt{q}-1}} \Sigma_1(g_1)\Sigma_2(g_2)\dots\Sigma_\mathtt{q}(g_\mathtt{q})\ket{\psi}^{\otimes n^{\mathtt{q}-1}},
\end{align}
where $\Sigma_\mathtt{k}(g_\mathtt{k})$ are twist operators for the permutation action of \(g_\mathtt{k}\) on indices for $A_\mathtt{k}$ of density matrices. In particular, for $\mathtt{q}=3$ tri-partite subsystems $A,B,C$, we contract indices of $n^2$ density matrices on each subsystem with the following permutation
\begin{align}
g_{A}&=(1,2,\dots,n)(n+1,n+2,\dots,2n)\cdots(n^2-n+1,n^2-n+2,\dots, n^2),\\
g_{B}&=(1,n+1,\dots,n^2-n+1)(2,n+2,\dots,n^2-n+2)\cdots(n,2n,\dots, n^2),\\
g_{C}&=(1)(2)\cdots(n^2).
\end{align}

We express the index structure of the density matrix $\rho=\ket{\psi}\bra{\psi}$ as
\begin{align}
\rho_{\alpha\beta\gamma}^{\alpha'\beta'\gamma'}=\left(\vert \psi\rangle\right)_{\alpha,\beta,\gamma}\left(\langle \psi\vert\right)^{\alpha',\beta',\gamma'},
\end{align}
where $(\alpha,\alpha'),(\beta,\beta'),(\gamma,\gamma')$ are indices of subsystems $A,B,C$, respectively. Then, the replica partition function $Z^{(\mathtt{3})}_n$ is given by the following contraction
\begin{align}
\label{eq:Zq3n}
Z^{(\mathtt{3})}_n=\prod_{i=1}^{n^2}\delta^{\alpha_{g_A(i)}}_{\alpha'_i}\delta^{\beta_{g_B(i)}}_{\beta'_i}\delta^{\gamma_{g_C(i)}}_{\gamma'_i}\prod_{j=1}^{n^2} \rho_{\alpha_j\beta_j\gamma_j}^{\alpha'_j\beta'_j\gamma'_j},
\end{align}
where examples of the permutation are
\begin{align}
g_A(1)=2,g_A(n)=1,g_B(1)=n+1,g_B(n)=2n,g_C(1)=1,g_C(n)=n.
\end{align}
In particular, the replica partition function $Z^{(\mathtt{3})}_2$ with $\mathtt{q}=3,n=2$ is given by
\begin{align}
\label{eq:Zq3n2}
Z^{(\mathtt{3})}_2=\rho_{\alpha_1\beta_1\gamma_1}^{\alpha_2\beta_3\gamma_1}\rho_{\alpha_2\beta_2\gamma_2}^{\alpha_1\beta_4\gamma_2}\rho_{\alpha_3\beta_3\gamma_3}^{\alpha_4\beta_1\gamma_3}\rho_{\alpha_4\beta_4\gamma_4}^{\alpha_3\beta_2\gamma_4}.
\end{align}

For $\mathtt{q}=3$ tri-partite subsystems $A,B$, and $C$, genuine Rényi multi-entropy $\text{GM}^{(\mathtt{3})}_n(A:B:C)$ is defined as \cite{Harper:2024ker, Liu:2024ulq, Iizuka:2025ioc}
\begin{align}
\label{q3genuinemulti}
\begin{split}
\text{GM}^{(\mathtt{3})}_n(A:B:C) &:=S^{(\mathtt{3})}_n(A:B:C)   -\frac{1}{2}\left(S_n^{(\mathtt{2})}(AB:C)+S_n^{(\mathtt{2})}(BC:A)+S_n^{(\mathtt{2})}(CA:B)\right),
 \end{split}
\end{align}
where $S_n^{(\mathtt{2})}$ is the usual Rényi entropy for $\mathtt{q}=2$ bi-partite subsystems. As computed by \cite{Penington:2022dhr}, one can confirm that $\text{GM}^{(\mathtt{3})}_n(A:B:C)$ vanishes for triangle states
\begin{align}
\ket{\Delta}_{ABC}= \ket{\psi_1}_{A_LB_R}\otimes\ket{\psi_2}_{B_LC_R}\otimes\ket{\psi_3}_{C_LA_R} \,,
\label{eq:triangle}
\end{align}
where we decompose the Hilbert space of each subsystem, such as
\begin{align}
\mathcal{H}_A=\mathcal{H}_{A_L}\otimes\mathcal{H}_{A_R},\mathcal{H}_B=\mathcal{H}_{B_L}\otimes\mathcal{H}_{B_R},\mathcal{H}_C=\mathcal{H}_{C_L}\otimes\mathcal{H}_{C_R}.
\end{align}
Since the multi-entropy is additive under layer decomposition of pure states, $\text{GM}^{(\mathtt{3})}_n(A:B:C)$ also vanishes for the tensor products of triangle states. In this sense, a nonzero value of $\text{GM}^{(\mathtt{3})}_n(A:B:C)$ indicates the presence of genuinely tri-partite entanglement, which is not present in triangle states.

\subsection{Genuine Rényi multi-entropy of the GHZ state}
\label{subsec:GHZstate}

As a computable example, consider the $d$-dimensional GHZ state in tri-partite subsystems $A,B,$ and $C$ as follows:
\begin{align}
\ket{\text{GHZ}}=\frac{1}{\sqrt{d}}\sum_{l=0}^{d-1}\ket{lll}.
\label{GHZ}
\end{align}
Genuine Rényi multi-entropy of the $d$-dimensional GHZ state is given by \cite{Penington:2022dhr, Liu:2024ulq}
\begin{align}
\begin{split}
S^{(\mathtt{3})}_n(A:B:C)|_{\text{GHZ}}&=\frac{n+1}{n}\log d,\\
S_n^{(\mathtt{2})}(AB:C)|_{\text{GHZ}}&=S_n^{(\mathtt{2})}(BC:A)|_{\text{GHZ}}=S_n^{(\mathtt{2})}(CA:B)|_{\text{GHZ}}=\log d,\\
\text{GM}^{(\mathtt{3})}_n(A:B:C)|_{\text{GHZ}}&=\frac{2-n}{2n}\log d.
\end{split}
\label{GMGHZ1}
\end{align}
In particular, genuine $n=2$ Rényi multi-entropy of the GHZ state vanishes:
\begin{align}
\text{GM}^{(\mathtt{3})}_2(A:B:C)|_{\text{GHZ}}&=0.
\end{align}
Therefore, genuine Rényi multi-entropy can be zero even with genuinely multi-partite entanglement, such as the GHZ state, and cannot serve as a genuinely multi-partite entanglement measure for all types of genuinely multi-partite entanglement. Instead, genuine Rényi multi-entropy belongs to a class called \emph{signals} \cite{Gadde:2026msg}, which are local unitary invariant, additive under layer decomposition, and vanish for separable states. 

For the $d$-dimensional GHZ state \eqref{GHZ}, local purity $\mu$ is given by
\begin{align}
\mu=\Tr_A(\rho_A)^2=\Tr_B(\rho_B)^2=\Tr_C(\rho_C)^2=\frac{1}{d},
\end{align}
where reduced density matrices are defined by
\begin{align}
\rho_A:=\Tr_{BC} \rho, \;\;\;\rho_B:=\Tr_{CA} \rho,\;\;\;\rho_C:=\Tr_{AB} \rho.
\end{align}
Therefore, in terms of local purity $\mu$, genuine Rényi multi-entropy in eq. \eqref{GMGHZ1} is written as
\begin{align}
\text{GM}^{(\mathtt{3})}_n(A:B:C)|_{\text{GHZ}}&=\frac{n-2}{2n}\log \mu.
\label{GMGHZ}
\end{align}

One can consider the following generalization of the GHZ state for $N$-partite $d$-dimensional qudits 
\begin{align}
\ket{\text{GHZ}}=\frac{1}{\sqrt{d}}\sum_{l=0}^{d-1}|\underbrace{l\dots l}_{N}\rangle.
\label{NGHZ}
\end{align}
We divide the $N$-partite qudits into three subsystems $A,B,C$ such as
\begin{align}
\dim{\mathcal{H}_A}=d^{N_A}, \dim{\mathcal{H}_B}=d^{N_B}, \dim{\mathcal{H}_C}=d^{N_C},
\end{align}
where $N_A\ge1,N_B\ge1,N_C\ge1$, and $N=N_A+N_B+N_C$. Even for such a generalized GHZ state \eqref{NGHZ} with any choice of subsystems, the above formulas for the GHZ state \eqref{GHZ} hold true in the same way.
We will see that the genuine Rényi multi-entropy of fully symmetric Gaussian states behaves similarly to $\text{GM}^{(3)}_n(A:B:C)|_{\text{GHZ}}$.

\subsection{Formalism of the genuine Rényi multi-entropy for pure bosonic Gaussian states}
\label{subsec:MEGaussian}

Consider a general pure Gaussian state of $N$ bosonic modes described by the wavefunction
\begin{align}
    \label{eq:GaussPureState}
    \psi(\mathbf{q})=\left(\det\left(\frac{\mathbf{R}}{\pi}\right)\right)^{1/4}\exp\left[-\frac{1}{2}\mathbf{q}^{\top}\cdot \mathbf{W}\cdot \mathbf{q}\right]\,~,\quad \mathbf{W}=\mathbf{R}+i\mathbf{S}~,
\end{align}
where $\mathbf{q}=(q^{1},\ldots,q^{N})^{\top}$ is a $N$-dimensional vector in position space. Here, $\mathbf{R}=\mathrm{Re}(\mathbf{W})$ is a symmetric positive definite $N$-dimensional matrix, and $\mathbf{S}=\mathrm{Im}(\mathbf{W})$ is a symmetric $N$-dimensional matrix. Thus, $\mathbf{W}$ is a symmetric $N$-dimensional complex matrix $\mathbf{W}=\mathbf{W}^{\top}$. 

The density matrix operator $\rho(\mathbf{q},\mathbf{q}')$ for the pure Gaussian state~\eqref{eq:GaussPureState} is given by
\begin{align}
\label{eq:RedDensMat}
    \rho(\mathbf{q},\mathbf{q}') &=\left(\det\left(\frac{\mathbf{R}}{\pi}\right)\right)^{1/2} \exp\left[-\frac{1}{2}\left(\mathbf{q}^\top\cdot \mathbf{W}\cdot \mathbf{q}+{\mathbf{q}'}^\top\cdot \mathbf{W}^*\cdot \mathbf{q}'\right)\right]~,
\end{align}
which is Hermitian $\rho^{\ast}(\mathbf{q},\mathbf{q}')=\rho(\mathbf{q}',\mathbf{q})$. 

To divide the system into three subsystems $A,B,C$, we divide the $N$-dimensional vector $\mathbf{q}\in \mathbb{R}^N$ into three vectors $\mathbf{q}_A\in \mathbb{R}^{N_A},\mathbf{q}_B\in \mathbb{R}^{N_B},\mathbf{q}_C\in \mathbb{R}^{N_C}$, where $N_A+N_B+N_C=N$, and we express this decomposition as 
\begin{align}
    \mathbf{q}=
\begin{pmatrix}
\mathbf{q}_{A}\\
\mathbf{q}_{B}\\
\mathbf{q}_{C}
\end{pmatrix}.
\end{align}
To compute the replica partition function $Z^{(3)}_n$ for tri-partite entanglement between these subsystems, we need to evaluate the contraction such as eq.~\eqref{eq:Zq3n}. For continuous variables, the Kronecker delta in eq.~\eqref{eq:Zq3n} should be replaced with the following integral
\begin{align}
    \rho_{\alpha_j\beta_j\gamma_j}^{\alpha'_j\beta'_j\gamma'_j}&\to\rho(\mathbf{q}^{(j)},\mathbf{q}'^{(j)}),\\
    \delta^{\alpha_{j}}_{\alpha'_i}&\to\int_{\mathbb{R}^{N_A}} \mathrm{d}\mathbf{q}_{A}^{(j)}\int_{\mathbb{R}^{N_A}} \mathrm{d}\mathbf{q}_{A}'^{(i)}\,\delta^{(N_A)}(\mathbf{q}^{(j)}_{A}-\mathbf{q}_{A}'^{(i)}),\\
    \delta^{\beta_{j}}_{\beta'_i}&\to\int_{\mathbb{R}^{N_B}} \mathrm{d}\mathbf{q}_{B}^{(j)}\int_{\mathbb{R}^{N_B}} \mathrm{d}\mathbf{q}_{B}'^{(i)}\,\delta^{(N_B)}(\mathbf{q}^{(j)}_{B}-\mathbf{q}_{B}'^{(i)}),\\
    \delta^{\gamma_{j}}_{\gamma'_i}&\to\int_{\mathbb{R}^{N_C}} \mathrm{d}\mathbf{q}_{C}^{(j)}\int_{\mathbb{R}^{N_C}} \mathrm{d}\mathbf{q}_{C}'^{(i)}\,\delta^{(N_C)}(\mathbf{q}^{(j)}_{C}-\mathbf{q}_{C}'^{(i)}),
\end{align}
where $i$ and $j$ represent replica indices for replica density matrices. For example, the replica partition function $Z^{(3)}_2$ \eqref{eq:Zq3n2} for $n=2$ is given by
\begin{align}
\begin{split}
Z^{(3)}_2&=\int_{\mathbb{R}^{N}} \mathrm{d}\mathbf{q}^{(1)}\int_{\mathbb{R}^{N}} \mathrm{d}\mathbf{q}^{(2)}\int_{\mathbb{R}^{N}} \mathrm{d}\mathbf{q}^{(3)}\int_{\mathbb{R}^{N}} \mathrm{d}\mathbf{q}^{(4)}\\
&\times\rho((\mathbf{q}_A^{(1)},\mathbf{q}_B^{(1)},\mathbf{q}_C^{(1)}),(\mathbf{q}_A^{(2)},\mathbf{q}_B^{(3)},\mathbf{q}_C^{(1)}))\rho((\mathbf{q}_A^{(2)},\mathbf{q}_B^{(2)},\mathbf{q}_C^{(2)}),(\mathbf{q}_A^{(1)},\mathbf{q}_B^{(4)},\mathbf{q}_C^{(2)}))\\
&\times\rho((\mathbf{q}_A^{(3)},\mathbf{q}_B^{(3)},\mathbf{q}_C^{(3)}),(\mathbf{q}_A^{(4)},\mathbf{q}_B^{(1)},\mathbf{q}_C^{(3)}))\rho((\mathbf{q}_A^{(4)},\mathbf{q}_B^{(4)},\mathbf{q}_C^{(4)}),(\mathbf{q}_A^{(3)},\mathbf{q}_B^{(2)},\mathbf{q}_C^{(4)})).
\end{split}
\end{align}
To compute the replica partition functions, we use the Gaussian integral formula, such as
\begin{align}
    \int_{\mathbb{R}^{N}} \mathrm{d}\mathbf{q}\exp\left[-\frac{1}{2}\left(\mathbf{q}^\top\cdot\left( \mathbf{W}+\mathbf{W}^*\right)\cdot \mathbf{q}\right)\right]=\left(\det\left(\frac{\mathbf{R}}{\pi}\right)\right)^{-1/2},\;\;\;\mathbf{R}=\frac{\mathbf{W}+\mathbf{W}^*}{2}.
\end{align}
By using the Gaussian integral formula, as performed in \cite{Harper:2025uui}, one can express the replica partition functions as the determinant of certain sparse matrices. For example, computing the replica partition function $Z^{(\mathtt{q})}_{n}$ for general $N$ involves computing the determinant of a sparse matrix of dimensions $Nn^{\mathtt{q}-1}\times Nn^{\mathtt{q}-1}$.

\section{Genuine Rényi multi-entropy of fully symmetric Gaussian states}
\label{sec:MEGaussian}

In this section, we study the $\mathtt{q}=3$ genuine Rényi multi-entropy of fully symmetric Gaussian states. After explaining the fully symmetric Gaussian states, we explicitly show exact expressions of the genuine Rényi multi-entropy of fully symmetric Gaussian states with a few modes. Here, the fully symmetric Gaussian states are represented by one parameter $a$, where local purity $\mu$ is given by $\mu=1/a$. 

\subsection{Fully symmetric Gaussian states}
\label{subsec:FullyGaussStates}

We start to explain the definition of fully symmetric Gaussian states 
\cite{Adesso:2004fz}. A general pure Gaussian state of $N$ bosonic modes is represented by an $N\times N$ symmetric matrix $\mathbf{W}$, such as eq.~\eqref{eq:RedDensMat}. Let us call a fully symmetric Gaussian state \cite{Adesso:2004fz} represented by $\mathbf{W}$ such that all diagonal elements are $a\ge1$ and all off-diagonal elements are $e^-$, where
\begin{align}\label{offd}
e^-:=\frac{(a^2-1)(N-2)-\sqrt{a^2-1} \sqrt{(a^2-1)N^2+4(N-1)}}{2 a (N-1)}.
\end{align}
For example, $\mathbf{W}$ of the fully symmetric Gaussian state with $N=3$ modes is given by \cite{Adesso:2006ktx}
\begin{align}
\label{eq:Wmat3mode}
\mathbf{W}=\begin{pmatrix}
    a & \frac{a^2-1-\sqrt{a^2-1} \sqrt{9 a^2-1}}{4 a}& \frac{a^2-1-\sqrt{a^2-1} \sqrt{9 a^2-1}}{4 a} \\ \frac{a^2-1-\sqrt{a^2-1} \sqrt{9 a^2-1}}{4 a} & a & \frac{a^2-1-\sqrt{a^2-1} \sqrt{9 a^2-1}}{4 a}\\
    \frac{a^2-1-\sqrt{a^2-1} \sqrt{9 a^2-1}}{4 a}& \frac{a^2-1-\sqrt{a^2-1} \sqrt{9 a^2-1}}{4 a}& a
   \end{pmatrix}.
\end{align}
By choosing the off-diagonal elements as eq.~\eqref{offd}, one can check that \emph{local} purity $\mu$ is given by
\begin{align}\label{LocalPurity}
\mu=\Tr_{q^1}\left(\Tr_{\overline{q^1}}\rho\right)^2=\Tr_{q^2}\left(\Tr_{\overline{q^2}}\rho\right)^2=\dots=\Tr_{q^N}\left(\Tr_{\overline{q^N}}\rho\right)^2=\frac{1}{a}.
\end{align}
Henceforth, for simplicity, we use the notation $\Tr$ to represent the integral of  continuous variables, for instance, 
\begin{align}
    \Tr\rho=\int_{\mathbb{R}^{N}} \mathrm{d}\mathbf{q}\int_{\mathbb{R}^{N}} \mathrm{d}\mathbf{q}'\,\rho(\mathbf{q},\mathbf{q}')\delta^{(N)}(\mathbf{q}-\mathbf{q}').
\end{align}
The fully symmetric state belongs to the class of continuous-variable (CV) GHZ-type states, and in the limit of infinite squeezing $a\to\infty$, these states approach the proper CV GHZ state \cite{vanLoock:2002ydf}. 

Fully symmetric Gaussian states can be realized as ground states of quadratic (free) bosonic Hamiltonians that are permutation symmetric. Conversely, any ground state of a permutation-symmetric quadratic Hamiltonian has a fully symmetric covariance matrix. For $N$ bosonic modes with canonical coordinates $\mathbf{q}=(q^{1},\ldots,q^{N})^{\top}$ and momenta $\mathbf{p}=(p_{1},\ldots,p_{N})^{\top}$, the (canonical) quadratic Hamiltonian
\begin{align}
    \label{eq:FreeHamSymm}
   \mathbf{H}=\frac{1}{2}\left(\mathbf{p}^{\top}\cdot\mathbf{W}^{-1}\cdot\mathbf{p}+\mathbf{q}^{\top}\cdot \mathbf{W}\cdot\mathbf{q}\right)~,
\end{align}
with
\begin{align}
    \label{eq:Rmat}
    \mathbf{W}=\alpha\mathbb{I}_{N}+e^{-}\mathbf{J}_{N}~,\quad~\alpha:=(a-e^{-})~, 
\end{align}
\begin{align}
    \label{eq:Rinvmat}
     \mathbf{W}^{-1}=\frac{1}{\alpha}\mathbb{I}_{N}-\frac{e^{-}}{\alpha(\alpha+Ne^{-})}\mathbf{J}_{N}~,\quad 
\end{align}
where $\mathbb{I}_{N}$ is the $N\times N$ identity matrix and $\mathbf{J}_{N}$ is the $N\times N$ matrix with entries $(\mathbf{J}_{N})_{ij}=1\quad\forall\,i,j=1,\ldots,N$, has the fully symmetric Gaussian state with matrix $\mathbf{W}$ defined according to the definition before~\eqref{offd} as its ground state. Note that the wave function \eqref{eq:GaussPureState} is the ground state of Hamiltonian \eqref{eq:FreeHamSymm} only if $\mathbf{W}$ is a real matrix, like~\eqref{offd}. We provide additional details in App.~\ref{app2}.

\subsection{Genuine Rényi multi-entropy of fully symmetric Gaussian states}
\label{subsec:ResultsRMEFullySGauss}

Once $\mathbf{W}$ of the fully symmetric state is given, we can systematically compute the genuine Rényi multi-entropy, although the computational cost increases when $n$ or $N$ is large. Our calculation results are summarized in Table \ref{table:GM}. Here, we denote $N_A,N_B,N_C$ as the number of modes in subsystems $A,B,C$, respectively.

\begin{table}[t]
\centering
\resizebox{\linewidth}{!}{
\begin{tabular}{|c|c|c|c|}
\hline
  &$\text{GM}^{(\mathtt{3})}_2$&    $\text{GM}^{(\mathtt{3})}_3$ & $\text{GM}^{(\mathtt{3})}_4$ \\
\hline
\begin{tabular}{c}
$N=3$ modes\\
$(N_A=N_B=N_C=1)$
\end{tabular}
& $0$ & $\frac{1}{12} \log \left(\frac{\left(9 a^2-1\right)^2}{\left(3 a^2+1\right)^3}\right)$ 
& $\frac{1}{4} \log \left(\frac{2 a}{a^2+1}\right)$\\
\hline
 \begin{tabular}{c}
$N=4$ modes\\
$(N_A=2,N_B=N_C=1)$
\end{tabular}
 & 0  & 
 $\frac{1}{6} \log \left(\frac{5 a^2-1}{a(3 a^2+1)}\right)$&
  $\frac{1}{24} \log \left(\frac{\left(7 a^2-1\right)^4}{3 \left(a^2+1\right)^4 \left(2
   a^2+1\right)^2 \left(4 a^2-1\right)}\right)$\\
\hline
\begin{tabular}{c}
$N=5$ modes\\
$(N_A=3,N_B=N_C=1)$
\end{tabular}
 & 0  & $\frac{1}{12} \log \left(\frac{\left(21 a^2-5\right)^2}{2 \left(3 a^2+1\right)^2 \left(9
   a^2-1\right)}\right)$
 &
  $\frac{1}{24} \log \left(\frac{2 \left(5 a^2-1\right)^4}{\left(a^2+1\right)^4 \left(3
   a^2-1\right) \left(3 a^2+1\right)^2}\right)$\\
\hline
\begin{tabular}{c}
$N=5$ modes\\
$(N_A=N_B=2,N_C=1)$
\end{tabular}
 & 0  & $\frac{1}{12} \log \left(\frac{64 \left(3 a^2-1\right)^2}{\left(9 a^2-1\right)^2 \left(3
   a^2+1\right)}\right)$
 &
  $\frac{1}{12} \log \left(\frac{2 \left(5a^2-1\right)^2}{a\left(3 a^2+1\right)^2
   \left(a^2+1\right)}\right)$
 \\
\hline
\begin{tabular}{c}
$N=6$ modes\\
$(N_A=4,N_B=N_C=1)$
\end{tabular}
 & 0  & $\frac{1}{12} \log \left(\frac{\left(27 a^2-7\right)^2}{5 \left(3 a^2+1\right)^2 \left(6
   a^2-1\right)}\right)$
 &
  $\frac{1}{24} \log \left(\frac{\left(13 a^2-3\right)^4}{5 \left(a^2+1\right)^4 \left(4
   a^2+1\right)^2 \left(8 a^2-3\right)}\right)$
 \\
\hline
\begin{tabular}{c}
$N=6$ modes\\
$(N_A=3,N_B=2,N_C=1)$
\end{tabular}
 & 0  & $\frac{1}{12} \log \left(\frac{\left(33 a^2-13\right)^2}{\left(3 a^2+1\right) \left(6
   a^2-1\right) \left(27 a^2-7\right)}\right)$
 &
  $\frac{1}{24} \log \left(\frac{\left(7 a^2-2\right)^2 \left(13 a^2-3\right)^2 \left(17
   a^2-7\right)^2}{a^2 \left(a^2+1\right)^2 \left(4 a^2+1\right)^2 \left(8 a^2-3\right)
   \left(9 a^2-4\right) \left(9 a^2+1\right)^2}\right)$
 \\
\hline
\begin{tabular}{c}
$N=6$ modes\\
$(N_A=N_B=N_C=2)$
\end{tabular}
 & 0  & $\frac{1}{12} \log \left(\frac{5 \left(9 a^2-4\right)^2}{\left(6 a^2-1\right)^3}\right)$
 &
  $\frac{1}{8} \log \left(\frac{5 \left(8 a^2-3\right)}{\left(4 a^2+1\right)^2}\right)$
 \\
\hline
\begin{tabular}{c}
$N=7$ modes\\
$(N_A=5,N_B=N_C=1)$
\end{tabular}
 & 0  & $\frac{1}{12} \log \left(\frac{(55a^{4}-26a^{2}+3)^{2}}{(3a^{2}+1)^{2}(5a^{2}-1)^{3}}\right)$
 &
  $\frac{1}{24}\log\left(\frac{2^{6}(5a^{2}-2)\left(4a^{4}+3a^{2}-1\right)^{4}}{3(1+a^{2})^{8}(25a^{4}-5a^{2}-2)^{2}}\right)$
 \\
\hline
\begin{tabular}{c}
$N=8$ modes\\
$(N_A=6,N_B=N_C=1)$
\end{tabular}
 & 0  & $\frac{1}{12} \log \left(\frac{(351a^{4}-177a^{2}+22)^{2}}{7(9a^{2}-2)^{3}(3a^{2}+1)^{2}}\right)$
 &
  $\frac{1}{24}\log\left(\frac{(19a^2-5)^4}{7(1+a^{2})^{4}(1+6a^{2})^{2}(12a^2-5)}\right)$
 \\
\hline
\end{tabular}
}
\caption{Exact values of the genuine Rényi multi-entropy $\text{GM}^{(3)}_n$ for fully symmetric Gaussian states. Here, $a$ is the inverse of the local purity, as shown in eq.~\eqref{LocalPurity}. The number of modes in each subsystem $A,B,C$ is denoted by $N_A,N_B,N_C$, respectively, and $N=N_A+N_B+N_C$ is the total number of modes.}
\label{table:GM}
\end{table}

From Table \ref{table:GM}, we can observe the following properties of the genuine Rényi multi-entropy of fully symmetric Gaussian states.
\begin{itemize}
    \item If $a=1$, off-diagonal elements $e^-$ \eqref{offd} vanish as $e^-=0$. This means that there is no entanglement in the fully symmetric Gaussian state if $a=1$. Therefore, the genuine Rényi multi-entropy in Table \ref{table:GM} becomes zero if $a=1$.
    \item If $n=2$, the genuine Rényi multi-entropy of the fully symmetric Gaussian states is zero for any values of $a$ and $N$. This property of $\text{GM}^{(\mathtt{3})}_2$ for Gaussian states was first numerically observed in \cite{Harper:2025uui} for the ground state of a harmonic chain. In Section \ref{sec:ProogGMn2}, we analytically show $\text{GM}^{(\mathtt{3})}_2=0$ for any pure bosonic Gaussian state, even with correlations between position and momentum operators.
    \item In the limit of infinite squeezing $a\to\infty$, the fully symmetric Gaussian states approach the proper CV GHZ state. Hence, at $a\to\infty$, $\text{GM}^{(3)}_n$ is expected to be the same expression as the GHZ state \eqref{GMGHZ} with $\mu=1/a$. Indeed, $\text{GM}^{(3)}_n$ in Table \ref{table:GM} has the following asymptotic behavior at $a\to\infty$
    \begin{align}\label{AsymB}
    \text{GM}^{(\mathtt{3})}_n\sim \frac{2-n}{2n}\log a
\;\;\;(a\to\infty).    
    \end{align}
    Based on the relationship between the fully symmetric Gaussian states and the proper CV GHZ state, we \emph{conjecture} that the asymptotic behavior \eqref{AsymB} is valid for the fully symmetric Gaussian states with any $n$ and $N$.
    \item In the limit of \emph{large} squeezing $a\gg 1$, fixing $N_{B}=N_{C}=1$, the behavior of $\textrm{GM}_{3}^{(3)}$ depends on $N_{A}$ according to 
    \begin{align}
    \label{eq:FormualGq3n3NB1NC1}
    \begin{split}
   \text{GM}^{(\mathtt{3})}_{3}(a,N_A)\sim& -\frac{1}{6}\log(a)+\frac{1}{12}\log\left(\frac{2\left((4N_{A}(N_{A}+1)+1\right)}{3N_{A}(N_{A}+1)}\right)\\
   &-\left(\frac{2+3N_{A}+6N_{A}^2}{36(N_{A}(2N_{A}+1))}\right)\frac{1}{a^{2}}+O\left(\frac{1}{a^{4}}\right)~,\quad (a\gg 1~,N_{B}=N_{C}=1)~.
   \end{split}
\end{align}
For $n\geq 4$, it is difficult to find a similar closed-form expression.
\item In the limit of \emph{small} squeezing $0<(a-1)\ll 1$, fixing $N_{B}=N_{C}=1$, the behavior of $\textrm{GM}_{3}^{(\mathtt{3})}$ depends on $N_{A}$ according to 
    \begin{align}
    \label{eq:FormualGq3n3NB1NC1Smalla}
    \begin{split}
   \text{GM}^{(\mathtt{3})}_{3}(a,N_A)\sim& -\frac{3N_{A}(N_{A}+2)}{16(N_{A}+1)^{2}}(a-1)^{2}~,\quad (1\gg a-1>0~,N_{B}=N_{C}=1)~.
   \end{split}
\end{align}
For $n\geq 4$, it is difficult to find a similar closed-form expression.
\end{itemize}

\section{Proof of \texorpdfstring{$\text{GM}^{(\mathtt{3})}_2(A:B:C)=0$}{GMn20} for general pure bosonic Gaussian states}\label{sec:ProogGMn2}

In this section, we provide an analytic proof of the general vanishing of the genuine multi-entropy for $\mathtt{q}=3,n=2$, $\text{GM}^{(\mathtt{3})}_{2}=0$ for general pure bosonic Gaussian states, even with correlation between position and momentum operators.

From the density matrix $\rho(\mathbf{q},\mathbf{q}')$ \eqref{eq:RedDensMat}, we define the reduced density matrix $\rho_{AB}(\mathbf{q}_{AB},\mathbf{q}'_{AB})$ on $AB$ by tracing out the degrees of freedom of $C$
\begin{align}
\label{eq:ReducedDM}
   \rho_{AB}(\mathbf{q}_{AB},\mathbf{q}'_{AB})=\Tr_C\rho=\int_{\mathbb{R}^{N_C}} \mathrm{d}\mathbf{q}_{C}\int_{\mathbb{R}^{N_C}} \mathrm{d}\mathbf{q}'_C\,\rho(\mathbf{q},\mathbf{q}')\delta^{(N_C)}(\mathbf{q}_{C}-\mathbf{q}_{C}'),
\end{align}
where 
\begin{align}
    \mathbf{q}_{AB}=
\begin{pmatrix}
\mathbf{q}_{A}\\
\mathbf{q}_{B}
\end{pmatrix}.
\end{align}
Like the convention in \cite{Harper:2025uui}, we decompose the matrix $\mathbf{W}$ in the following way
\begin{align}
\label{eq:Wmat}
    \mathbf{W}&=\begin{pmatrix}
     \mathbf{A}_{ab} & \mathbf{B}_{a\beta} \\ \mathbf{B}^{\top}_{\alpha b} & \mathbf{C}_{\alpha\beta}
 \end{pmatrix}~,\quad \mathbf{W}^{\ast}=\begin{pmatrix}
    \mathbf{A}^*_{ab} & \mathbf{B}^*_{a\beta} \\ \mathbf{B}^{\dagger}_{\alpha b} & \mathbf{C}^*_{\alpha\beta}
   \end{pmatrix}~,
\end{align}
where Latin indices run over the degrees of freedom of $AB$, and Greek indices run over the degrees of freedom of $C$. That means, $\mathbf{A}$ acts on $AB$, $\mathbf{C}$ acts on $C$, and $\mathbf{B}$ couples $AB$ to $C$, where $\mathbf{A}$ and $\mathbf{C}$ are symmetric matrices $\mathbf{A}^\top=\mathbf{A}, \mathbf{C}^\top=\mathbf{C}$. Each individual block is complex and can be decomposed as $\mathbf{A}=\mathrm{Re}(\mathbf{A})+i\mathrm{Im}(\mathbf{A})$, but we keep the notation compact. Performing the Gaussian integral in eq.~\eqref{eq:ReducedDM} for $\mathbf{q}_{C}$ yields
\begin{align}
    \label{eq:RedDMatAB}
    \begin{split}
   \rho_{AB}(\mathbf{q}_{AB},\mathbf{q}'_{AB})=\det\left(\frac{\mathrm{Re}(\mathbf{K}+\mathbf{L})}{\pi}\right)^{1/2}\exp\Big[&-\frac{1}{2}\mathbf{q}_{AB}^{\top}\cdot \mathbf{K} \cdot \mathbf{q}_{AB}-\frac{1}{2}\mathbf{q}_{AB}^{'\top}\cdot \mathbf{K}^{\ast} \cdot \mathbf{q}'_{AB}\\
   &-\frac{1}{2}\mathbf{q}_{AB}^{\top}\cdot \mathbf{L}\cdot \mathbf{q}'_{AB}-\frac{1}{2}\mathbf{q}_{AB}^{'\top}\cdot \mathbf{L}^*\cdot \mathbf{q}_{AB}\Big]~,
   \end{split}
\end{align}
where 
\begin{align}
    \label{eq:MatricesRedDMatAB}
    \mathbf{K}=\mathbf{A}-\frac{1}{2}\mathbf{B}(\mathbf{\Sigma}_{C})^{-1}\mathbf{B}^{\top}~,\,\quad \mathbf{L}=-\frac{1}{2}\mathbf{B}(\mathbf{\Sigma}_{{C}})^{-1}\mathbf{B}^{\dagger}~,\,\quad \mathbf{\Sigma}_{{C}}=\frac{1}{2}\left(\mathbf{C}+\mathbf{C}^{\ast}\right)~.
\end{align}

By using the reduced density matrix $\rho_{AB}$, we can evaluate the $n=2$ Rényi entropy $S_2^{(2)}$ as follows
\begin{align}
    S_2^{(2)}(BC:A)&=-\log \mathcal E_{A},\;\;\; \mathcal E_{A}:=\Tr_A(\Tr_B\rho_{\mathrm{AB}})^2,\\
    S_2^{(2)}(CA:B)&=-\log \mathcal E_{B},\;\;\; \mathcal E_{B}:=\Tr_B(\Tr_A\rho_{\mathrm{AB}})^2,\\
    S_2^{(2)}(AB:C)&=-\log \mathcal E_{C},\;\;\; \mathcal E_{C}:=\Tr_{AB}(\rho_{\mathrm{AB}})^2.
\end{align}
We can also compute the replica partition function $Z^{(\mathtt{3})}_2$ \eqref{eq:Zq3n2} because the density matrices with the contraction of indices for $C$ in eq.~\eqref{eq:Zq3n2} correspond to $\rho_{AB}$. The genuine multi-entropy $\text{GM}^{(\mathtt{3})}_{2}$ for $\mathtt{q}=3,n=2$ is given by
\begin{align}
\begin{split}
    \text{GM}^{(3)}_2(A:B:C) &=S^{(3)}_2(A:B:C)   -\frac{1}{2}\left(S_2^{(2)}(AB:C)+S_2^{(2)}(BC:A)+S_2^{(2)}(CA:B)\right)\\
    &=-\frac{1}{2}\log Z^{(3)}_2-\frac{1}{2}\left(-\log \mathcal E_{C}-\log \mathcal E_{A}-\log \mathcal E_{B}\right)=-\frac{1}{2}\log \frac{Z^{(3)}_2}{\mathcal E_{A} \mathcal E_{B} \mathcal E_{C}},
    \end{split}
\end{align}
where we used $Z^{(\mathtt{q})}_1=1$.
Therefore, we conclude that $\text{GM}^{(\mathtt{3})}_2(A:B:C)=0$ if $Z^{(\mathtt{3})}_2=\mathcal E_{A} \mathcal E_{B} \mathcal E_{C}$.

To prove $Z^{(\mathtt{3})}_2=\mathcal E_{A} \mathcal E_{B} \mathcal E_{C}$, we give matrix representation of the replica partition functions. Their representation is given by
\begin{align}
\begin{split}
    \mathcal E_{A}&=\Tr_A(\Tr_B\rho_{\mathrm{AB}})^2=\mathrm{det}\left(\mathrm{Re}(\mathbf{K}+\mathbf{L})\right)\left(\det \mathcal M_A\right)^{-1/2}\\
&=\mathrm{det}\left(\mathrm{Re}(\mathbf{K}+\mathbf{L})\right)\left(\det(M_{A1}+M_{A2})\det(M_{A1}- M_{A2})\right)^{-1/2},
\end{split}\\
\begin{split}
    \mathcal E_{B}&=\Tr_B(\Tr_A\rho_{\mathrm{AB}})^2=\mathrm{det}\left(\mathrm{Re}(\mathbf{K}+\mathbf{L})\right)\left(\det \mathcal M_B\right)^{-1/2}\\
&=\mathrm{det}\left(\mathrm{Re}(\mathbf{K}+\mathbf{L})\right)\left(\det(M_{B1}+M_{B2})\det(M_{B1}- M_{B2})\right)^{-1/2},
\end{split}\\
\begin{split}
    \mathcal E_{C}&=\Tr_{AB}(\rho_{\mathrm{AB}})^2=\mathrm{det}\left(\mathrm{Re}(\mathbf{K}+\mathbf{L})\right)\left(\det \mathcal M_C\right)^{-1/2}\\
&=\mathrm{det}\left(\mathrm{Re}(\mathbf{K}+\mathbf{L})\right)\left(\det(M_{C1}+M_{C2})\det(M_{C1}- M_{C2})\right)^{-1/2},
\end{split}\\
\begin{split}
    Z^{(\mathtt{3})}_2&=\left(\mathrm{det}\left(\mathrm{Re}(\mathbf{K}+\mathbf{L})\right)\right)^2\left(\det \mathcal M_M\right)^{-1/2}\\
&=\left(\mathrm{det}\left(\mathrm{Re}(\mathbf{K}+\mathbf{L})\right)\right)^2\\
&\times\left(\det(M_{M1}+M_{M2}+M_{M3}+M_{M4})\det(M_{M1}-M_{M2}+M_{M3}-M_{M4})\right)^{-1/2}\\
&\times\left(\det(M_{M1}+M_{M2}-M_{M3}-M_{M4})\det(M_{M1}-M_{M2}-M_{M3}+M_{M4})\right)^{-1/2},
\end{split}
\end{align}
where 
\begin{align}\label{Matrices}
\begin{split}
\mathcal M_A&=
\begin{pmatrix}
M_{A1} & M_{A2} \\
M_{A2} & M_{A1}  
 \end{pmatrix},\;\;\; 
 \mathcal M_B=
\begin{pmatrix}
M_{B1} & M_{B2} \\
M_{B2} & M_{B1}  
 \end{pmatrix}, \;\;\;
  \mathcal M_C=
\begin{pmatrix}
M_{C1} & M_{C2} \\
M_{C2} & M_{C1}  
 \end{pmatrix}, \\
 \mathcal M_M &=
 \begin{pmatrix}
 M_{M1} & M_{M2} & M_{M3} & M_{M4} \\
 M_{M2} & M_{M1} & M_{M4} & M_{M3} \\
 M_{M3} & M_{M4} & M_{M1} & M_{M2} \\
 M_{M4} & M_{M3} & M_{M2} & M_{M1}
 \end{pmatrix}.
 \end{split}
\end{align}
See Appendix \ref{app1} for more details. 

By using the results in Appendix \ref{app1}, one can confirm the following relations
\begin{align}
    M_{M1}+M_{M2}+M_{M3}+M_{M4}&=M_{A1}+M_{A2}=M_{B1}+M_{B2}=M_{C1}+M_{C2}
    =\mathrm{Re}(\mathbf{K}+\mathbf{L}),
    \\
    M_{M1}-M_{M2}+M_{M3}-M_{M4}&=M_{A1}-M_{A2},\\
M_{M1}+M_{M2}-M_{M3}-M_{M4}&=M_{B1}-M_{B2},\\
M_{M1}-M_{M2}-M_{M3}+M_{M4}&=M_{C1}-M_{C2}.
\end{align}
Thus, we obtain
\begin{align}
    \begin{split}Z^{(3)}_2&=\left(\mathrm{det}\left(\mathrm{Re}(\mathbf{K}+\mathbf{L})\right)\right)^{3/2} \left(\det(M_{M1}-M_{M2}+M_{M3}-M_{M4})\right)^{-1/2}\\
    &\times\left(\det(M_{M1}+M_{M2}-M_{M3}-M_{M4})\det(M_{M1}-M_{M2}-M_{M3}+M_{M4})\right)^{-1/2}\\
    &=\left(\mathrm{det}\left(\mathrm{Re}(\mathbf{K}+\mathbf{L})\right)\right)^{3}\left(\det(M_{A1}+M_{A2})\det(M_{A1}- M_{A2})\right)^{-1/2}\\
    &\times \left(\det(M_{B1}+M_{B2})\det(M_{B1}- M_{B2})\det(M_{C1}+M_{C2})\det(M_{C1}- M_{C2})\right)^{-1/2}\\
    &=\mathcal E_{A} \mathcal E_{B} \mathcal E_{C}.
    \end{split}
\end{align}
Therefore, we conclude $Z^{(\mathtt{3})}_2=\mathcal E_{A} \mathcal E_{B} \mathcal E_{C}$ for general pure bosonic Gaussian states, which proves $\text{GM}^{(\mathtt{3})}_2(A:B:C)=0$.


\section{The Gaussian Multi-Entropy Curve}
\label{sec:GaussianMECurve}

In this section, we discuss the multi-entropy analogue of the Page curve, that describes the unitary time evolution of entanglement entropy in black hole evaporation~\cite{Page:1993df,Page:1993wv}, in the context of fully symmetric Gaussian states. The multi-entropy curve of Haar-random states in the context of black hole evaporation has been studied in~\cite{Iizuka:2024pzm,Iizuka:2025ioc}. Here, we consider a tripartite quantum system whose state is fully-symmetric and Gaussian and thus described by a matrix $\mathbf{W}$ as discussed in Sec.~\ref{subsec:FullyGaussStates}. 

In the original paper \cite{Page:1993wv} on the Page curve, it is assumed that an evaporating black hole system can be described by a Haar-random state because an evaporating black hole is considered a chaotic object. Based on this assumption, the Page curve of black hole evaporation can be analyzed using the entanglement entropy of a Haar-random state \cite{Page:1993df} with a fixed total Hilbert space dimension, where varying the Hilbert space dimension of subsystem is regarded as the unitary time evolution.

For the black hole multi-entropy curve of a tripartite random state  in~\cite{Iizuka:2024pzm,Iizuka:2025ioc}, it is assumed that one subsystem represents the degrees of freedom of an evaporating black hole, while the other two subsystems represent the degrees of freedom of two Hawking radiation subsystems. Initially, the black hole will contain all the modes of the full system, and the radiation systems will contain none of them. At the end of the evolution and once the black hole has fully evaporated, the black hole subsystem will contain no modes, whereas the radiation subsystems will contain all modes, distributed equally between them.

In the same spirit, we study the (genuine) multi-entropy curve of fully symmetric Gaussian states. Note that, since fully symmetric Gaussian states are not random states, our setting is not suitable to describe black hole evaporation. One interpretation of our analysis involves calculating the multi-entropy when the Hilbert space dimension of subsystem is varied.

In our analysis, we consider a tripartite density matrix  $\rho(\mathbf{q},\mathbf{q}')$ of a fully symmetric Gaussian state with fixed total number of modes $N=N_{A}+N_{B}+N_{C}=24$. We assume that subsystem $A$ initially contains all the modes of the full system, and subsystems $B$ and $C$ contain none of them. Then, we suppose that the degrees of freedom of $A$ gradually decrease, while those of $B$ and $C$ increase under the condition $N_B=N_C$.
Finally, $A$ will contain no modes, whereas subsystems $B$ and $C$ will contain all modes, distributed equally among them. During each step, the number of modes in $A$ will decrease by $2$, while the number of modes in $B$ and $C$  will each increase by $1$. Thus, a natural parameter that describes the change of tripartite entanglement in our setting is the combination 
\begin{align}
    \label{eq:MultiEntPageTime}
    t=N_{B}+N_{C}~,
\end{align}
which in the present case has the range $t\in [0,24]$ and increases in discrete steps of $2$.

In the black hole multi-entropy curve \cite{Iizuka:2024pzm}, there are two important time scales. One is the Page time, which is defined by the time scale when the dimension of two subsystems for $\mathbf{q}=2$ entanglement entropy is the same. The other one is the multi-entropy time, which is defined by the time scale when the dimension of multi-partite subsystems for  multi-entropy is the same. In our setting, the Page time is given by
\begin{align}
    t_{Page}=\left(N_{B}+N_{C}\right)|_{N_B+N_C=N_A}=12,
\end{align}
and the multi-entropy time is given by
\begin{align}
    t_{ME}=\left(N_{B}+N_{C}\right)|_{N_A=N_B=N_C}=16.
\end{align}

Note that we use the symbol $t$ for the parameter \eqref{eq:MultiEntPageTime} as the analogy of the Page curve, but it is unclear whether this parameter can be interpreted as ``time". This interpretation is valid if there exists a Hamiltonian of the time evolution that  allows the modes of subsystems to change while keeping the whole system fixed at the fully symmetric Gaussian state. Even if this interpretation does not work, our analysis can be regarded as a calculation of the dependence of multi-entropy on subsystem size.

\begin{figure}[ht]
    \centering
    \includegraphics[width=0.45\textwidth]{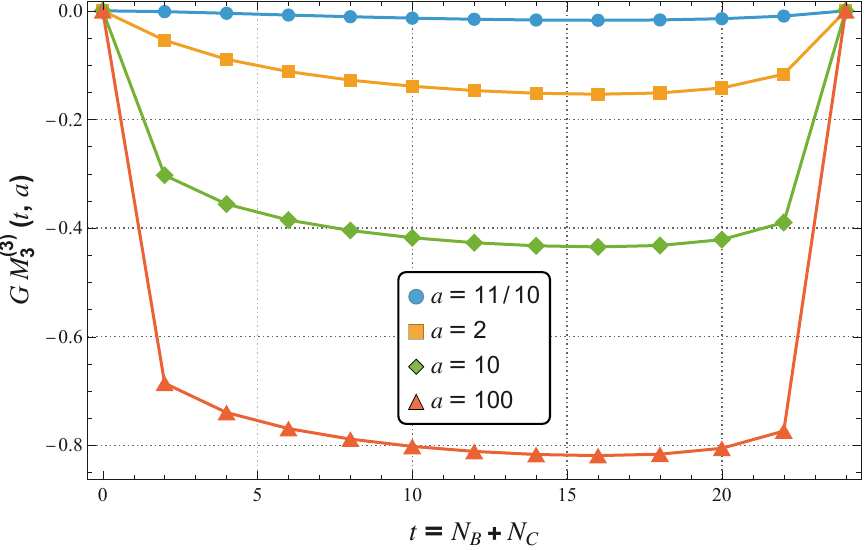}
     \caption{Rényi genuine multi-entropy curves $\textrm{GM}^{(3)}_{3}(t,a)$ for different values of $a=11/10,2,10,100$ and for $t=N_{B}+N_{C}\in[0,24]$, where  $N=N_A+N_B+N_C=24$ and $N_B=N_C$.}
\label{fig:PlotGMq3n3AllaPage}
\end{figure}

\begin{figure}[ht]
    \centering
    \begin{tabular}{cc}
        \includegraphics[width=0.45\textwidth]{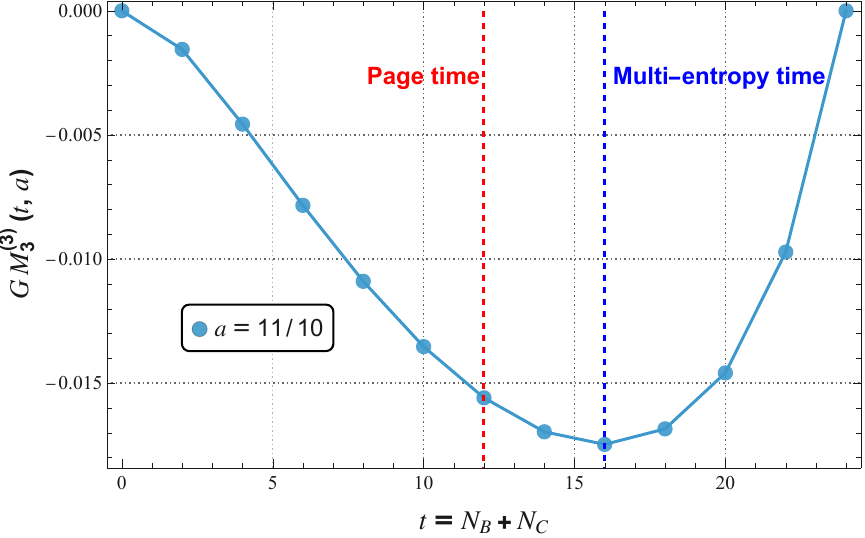} & 
        \includegraphics[width=0.45\textwidth]{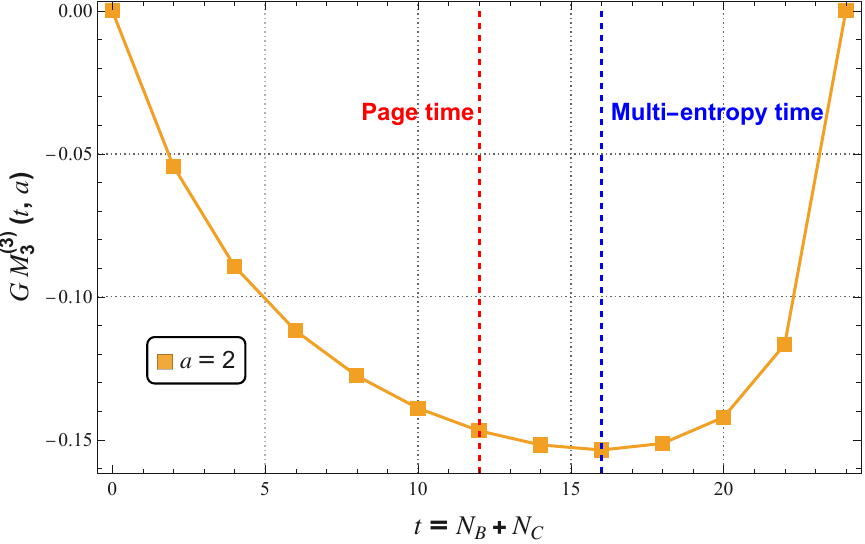} \\
      \includegraphics[width=0.45\textwidth]{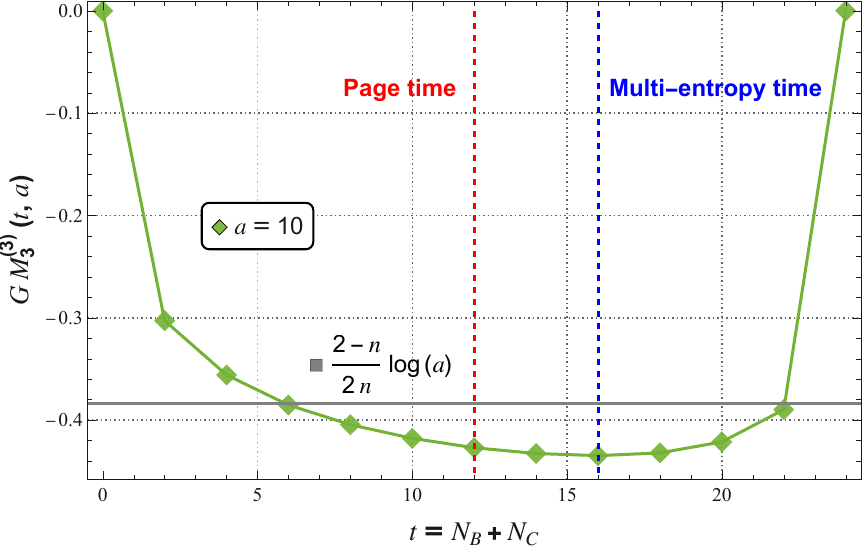}& 
        \includegraphics[width=0.45\textwidth]{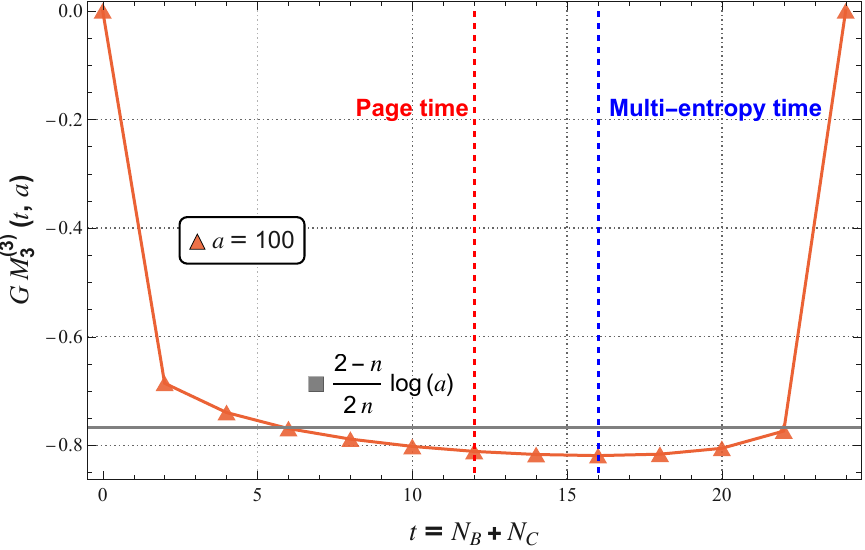}  \\
    \end{tabular}
     \caption{Individual $n=3$ Rényi genuine multi-entropy curves $\textrm{GM}^{(3)}_{3}(t,a)$ for different values of $a$ and for $t=N_{B}+N_{C}\in[0,24]$, where  $N=N_A+N_B+N_C=24$ and $N_B=N_C$. The vertical dashed lines represent the Page time (red) given by $t_{Page}=\left(N_{B}+N_{C}\right)|_{N_B+N_C=N_A}=12$ and the multi-entropy time (blue) given by $t_{ME}=\left(N_{B}+N_{C}\right)|_{N_A=N_B=N_C}=16$. The (gray) horizontal line in the plots for $a=10,100$ (bottom two) correspond to the behavior of the proper CV GHZ state~\eqref{GMGHZ} with $\mu=1/a$ and $n=3$.}
\label{fig:IndividualPlotsGMq3n3Page}
\end{figure}

Our interest is the $\mathtt{q}=3$ genuine $n$-th Rényi multi-entropy~\eqref{q3genuinemulti} as a function of $t$. For simplicity, we focus on $n=3$, which for these kind of states yields the first non-vanishing Rényi multi-entropy. Since fully-symmetric Gaussian states depend only on the local purity $1/a$, we consider different choices of this parameter and study their $t$-dependence. For a fixed $t=N_{B}+N_{C}$ with $N_{B}=N_{C}$ and $N_{A}=24-(N_{B}+N_{C})$, we compute $\textrm{GM}^{(3)}_{3}(a,t)$ for $t\in [0,24]$. The numerical results for $a=11/10,2,10,100$ can be seen in Fig.~\ref{fig:PlotGMq3n3AllaPage} and Fig.~\ref{fig:IndividualPlotsGMq3n3Page}.

From these figures, we note that the individual multi-entropy curves continue to decrease past the Page time $t_{Page}=\left(N_{B}+N_{C}\right)|_{N_B+N_C=N_A}=12$ until the multi-entropy time $t_{ME}=\left(N_{B}+N_{C}\right)|_{N_A=N_B=N_C}=16$, where they reach their minimum value at $t=t_{ME}$. After $t=t_{ME}$, they begin to increase, vanishing completely at $t=\left(N_{B}+N_{C}\right)|_{N_A=0}=24$. Thus, the process described here closely resembles the Page curve of an evaporating black hole, but the relevant scale when its behavior changes is $t_{ME}$, not $t_{Page}$.

\begin{figure}[ht]
    \centering
    \includegraphics[width=0.45\textwidth]{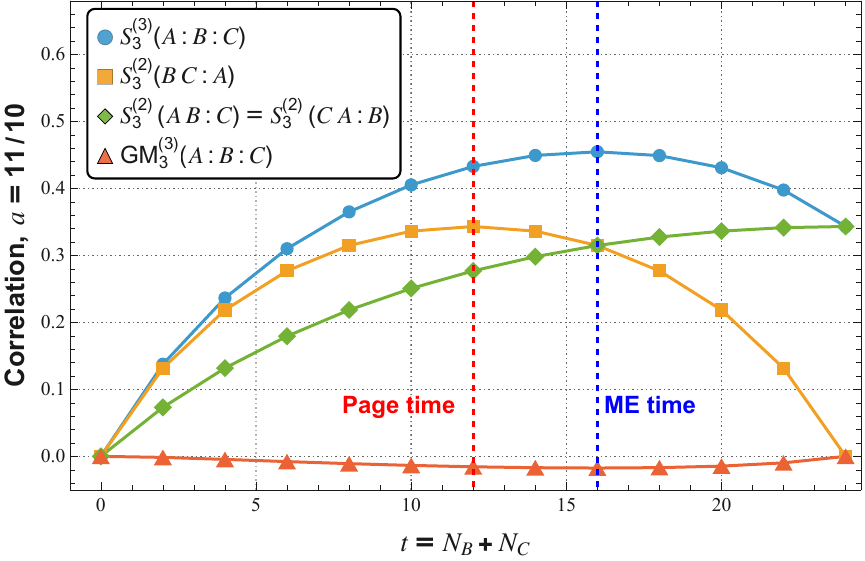}
     \caption{Comparison of $n=3$ Rényi entropies for different subsystems $S^{(2)}_{3}(AB:C)$, $S^{(2)}_{3}(BC:A)$, $S^{(2)}_{3}(CA:B)$, Rényi multi-entropy $S^{(3)}_{3}(A:B:C)$ and genuine Rényi multi-entropy $\textrm{GM}^{(3)}_{3}(A:B:C)$ with $a=11/10$ (shown in detail in the top-left plot of Fig.~\ref{fig:IndividualPlotsGMq3n3Page}) for $t=N_{B}+N_{C}\in[0,24]$, where $N=N_A+N_B+N_C=24$ and $N_B=N_C$. The vertical dashed lines represent the Page time (red) given by $t_{Page}=\left(N_{B}+N_{C}\right)|_{N_B+N_C=N_A}=12$ and the multi-entropy (ME) time (blue) given by $t_{ME}=\left(N_{B}+N_{C}\right)|_{N_A=N_B=N_C}=16$.}
\label{fig:PlotComparisonq3n3Pagea1p1}
\end{figure}

We can do a closer analysis of the individual contributions to the genuine multi-entropy $\textrm{GM}_{3}^{(3)}(A:B:C)$~\eqref{q3genuinemulti} to understand their relevant behavior at different $t$, namely at the Page time $t_{Page}$ and the multi-entropy time $t_{ME}$. Figure~\ref{fig:PlotComparisonq3n3Pagea1p1} shows the different contributions to the Rényi multi-entropy for $a=11/10$, where their behavior is the sharpest to compare.

From Figure~\ref{fig:PlotComparisonq3n3Pagea1p1}, we note that different contributions to the genuine Rényi multi-entropy $\textrm{GM}^{(\mathtt{3})}_{3}(A:B:C)$ behave differently at different $t$. For instance, the Rényi entropy $S_{3}^{(\mathtt{2})}(BC:A)$, which captures bi-partite entanglement properties between $A$ and $BC$, peaks at the Page time $t_{Page}$, whereas the Rényi multi-entropy $S_{3}^{(\mathtt{3})}(A:B:C)$ peaks instead at the multi-entropy time $t_{ME}$. The other contributions to the genuine multi-entropy, the Rényi entropies $S_{3}^{(\mathtt{2})}(AB:C)$=$S_{3}^{(\mathtt{2})}(CA:B)$
grow monotonically as $t$ increases. These results highlight the non-trivial properties of the Rényi genuine multi-entropy $\textrm{GM}^{(\mathtt{3})}_{3}(A:B:C)$ of fully symmetric Gaussian states with $N=24$, which appear to show a distinct change only at the multi-entropy time $t_{ME}$, even though $S_{3}^{(\mathtt{2})}(BC:A)$ changes its behavior at $t_{Page}$. This is supported by the analysis of the discrete first derivatives of all the Rényi entropy and multi-entropy curves.
These observations of fully symmetric Gaussian states are different from the genuine multi-entropy curve of Haar-random states with a large Hilbert space dimension~\cite{Iizuka:2025ioc}, where the genuine multi-entropy also changes its behavior at $t_{Page}$. This difference is likely due to the fact that we use a specific pure bosonic Gaussian state with a small value of $N$.

We can also compare our results with those of~\cite{Harper:2025uui}. Here, the authors study the Rényi multi-entropy for ground states of discretized $1+1$-dimensional free scalar conformal field theories (CFTs). In particular, the authors study quantities equivalent to the Rényi genuine multi-entropy in the context of the Gaussian ground state of a $N$-mode harmonic lattice model. Unlike the fully symmetric Gaussian states with a fixed local purity in our computations, 
the local purity of such a ground state depends on Hamiltonian parameters. Thus, it is interesting to contrast their behavior in similar setups. 

The results of~\cite{Harper:2025uui} for $\mathtt{q}=3$ are consistent with our findings: namely, the vanishing of the genuine Rényi multi-entropy for $n=2$, as well as a finite negative value for $n>2$. In particular, we can make a qualitative comparison between the middle plot of their Figure 5 and Fig.~\ref{fig:PlotGMq3n3AllaPage}, where the authors study the multi-entropy of three adjacent intervals on a circle with a fixed circumference $L$ as a function of the size of two of such adjacent intervals $\ell$, with the third having a size $L-2\ell$. The authors show the ``excess'' $\Delta S^{(3)}_{3}(A:B:C)$, a quantity equivalent to $\textrm{GM}^{(3)}_{3}(A:B:C)$, as a function of $\ell$, which displays a similar behavior to Fig.~\ref{fig:PlotGMq3n3AllaPage}. Although it is not possible to extract the precise behavior of $\Delta S^{(3)}_{3}(A:B:C)$ from their Figure, it behaves similarly to $\textrm{GM}^{(3)}_{3}(A:B:C)$, and the presence of a minimum can be inferred from the behavior at the edges of the plot, namely for $\ell = 1$ and $\ell = L-1$. Whether such a minimum coincides with the corresponding ``multi-entropy time'' $t_{\Delta S}=\ell|_{\ell=L-2\ell}=L/3$, is not evident from the plot alone, but cannot be discarded.


\section{Discussion}
\label{sec:Discussion}

In this manuscript we studied the tri-partite genuine Rényi multi-entropy GM$_{n}^{(\mathtt{3})}(A:B:C)$ of fully symmetric Gaussian states as a function of the squeezing parameter $a$ (inverse of the local purity $\mu$) for different total number of bosonic modes $N$ under different subsystem configurations $(N_{A},N_{B},N_{C})$. Restricting the parameter space of Gaussian states to a single parameter allowed us to find closed-form expressions for different replica index $n$.

A collection of exact values of the genuine Rényi multi-entropy GM$^{(\mathtt{3})}_{n}(A:B:C)$ for fully symmetric Gaussian states as a function of the squeezing parameter $a$ and for different system sizes can be found in Table~\ref{table:GM}. In particular, we found that as the squeezing parameter $a$ is taken to be very large, fully symmetric Gaussian states become insensitive to the details of the subsystem partitions, as well as to the total number of modes $N$. This is encapsulated by the asymptotic behavior shown in Eq.~\ref{AsymB}, which we conjecture to hold exactly in the limit $a\rightarrow \infty$ for any $n$ and $N$ and for any tri-partition. This is consistent with our expectation that fully symmetric Gaussian states approach a proper CV GHZ state in the limit $a\rightarrow \infty$, as mentioned in Sec.~\ref{subsec:FullyGaussStates}. Note that this expression is negative for $n>2$, and vanishes identically for $n=2$.

We provided two extensions of this result: Firstly, in Sec.~\ref{sec:ProogGMn2} we showed that the $n=2$ genuine Rényi multi-entropy GM$^{(\mathtt{3})}_{2}(A:B:C)$ of any generic tripartite pure bosonic $N$-mode Gaussian state vanishes identically due to a specific factorization property of the replica partition function $Z^{(\mathtt{3})}_{2}$. This factorization occurs due to the replica symmetry $\mathbb{Z}_2\times \mathbb{Z}_2$. As a consequence, this result also applies to the restricted subclass of fully symmetric Gaussian states introduced in Sec.~\ref{subsec:FullyGaussStates}. At the same time, this result provides an alternative perspective on the observation made in~\cite{Harper:2025uui} regarding the vanishing of the excess $\Delta S_{2}^{(\mathtt{3})}=0$ for subsystem tripartitions of the ground state of the one-dimensional harmonic chain. In that work, the authors state that this vanishing is due to the presence of the zero mode of the scalar field, present in non-compact scalar CFTs. This suggests that our squeezing parameter $a$ plays a role similar to the combination $m\delta$ in the lattice discretization of the $(1+1)$-dimensional massive scalar field, where $m$ is the mass of the scalar field and $\delta$ is the lattice spacing. Indeed, in Appendix~\ref{app2HarmChain} we note that the single-site purity in the thermodynamic limit of the Harmonic chain is primarily determined by the dimensionless combination $m\delta $ in two distinct limits~\eqref{eq:HarmChainPurityMass}. This suggests the following relation between the effects of the squeezing parameter of the fully symmetric Gaussian states $a$, and the combination $m\delta$ in the Harmonic chain on the purity of the single-site/local subsystem
\begin{align}
    \label{eq:ComparisonLocalPurity2}
    \begin{split}
       &m\delta\ll 1~\quad\longleftrightarrow \quad a\gg 1~,\\
        &m\delta\gg 1~\quad\longleftrightarrow \quad a\simeq 1~.
        \end{split}
\end{align}
Both a small mass $m\delta$ in the local chain and a large squeezing reduce the single-site/local purity, and conversely, large $m\delta$ and a squeezing parameter close to $1$ make the single-sites/local reduced states nearly pure. However, in general the harmonic chain's single-site purity is controlled by an integral over a $k-$dependent band and exhibits infrared divergences as $m\delta \rightarrow 0$. At the same time, the squeezing parameter $a$ cannot be directly interpreted as an infrared divergence, unlike $m\delta$. Nevertheless, the connection between $m\delta$ and $a$ may go beyond the fact that they reduce the local purity in the limit $m\delta\ll 1~\leftrightarrow a\gg 1$. In~\cite{Mollabashi:2024gik}, the authors discuss the preparation of the vacuum of a Klein-Gordon theory using a Gaussian/linear-optical network, and show that the squeezing required to reproduce the zero mode $k=0$, diverges logaritmically in the massless limit $m\rightarrow 0$. As a consequence, the massless limit of the non-compact scalar can be viewed as an infinite-squeezing limit of the zero-mode, requiring an infinite amount of energy to prepare. This operational fact about the CFT vacuum appears to be closely related to the identification~\eqref{eq:ComparisonLocalPurity2} and may reflect the common Gaussian mechanism that a soft zero-mode requires an arbitrarily large squeezing.\footnote{We thank Ali Mollabashi for pointing this out to us.}

Secondly, for $n=3$ we found closed-form expressions for $\textrm{GM}_{3}^{(\mathtt{3})}$ in the limits of large $a$~\eqref{eq:FormualGq3n3NB1NC1} and small $a$~\eqref{eq:FormualGq3n3NB1NC1Smalla} for fixed $N_{B}=N_{C}=1$ as a function of $N_{A}$. In particular, ~\eqref{eq:FormualGq3n3NB1NC1} shows the subleading behavior of ~\eqref{AsymB} for $n=2$ in powers of $a$ with coefficients depending only on $N_{A}$. This shows that the leading term in this expansion is always insensitive to details about subsystem sizes, and the subsystem size dependence only appears in the constant and subleading terms. This expression allows us to obtain an expectation for the $N\sim N_{A}\rightarrow \infty$ behavior of $\textrm{GM}_{3}^{(\mathtt{3})}$ subject to the condition $N_{B}=N_{C}=1$, given by
\begin{align}
    \label{eq:FormualGq3n3NB1NC1LargeaLargeNA}
    \begin{split}
   \text{GM}^{(\mathtt{3})}_{3}(a,N_A)\Big\vert^{(a\gg 1~,N_{B}=N_{C}=1)}_{N_{A}\rightarrow \infty}\rightarrow -\frac{1}{6}\log(a)+\frac{1}{12}\log\left(\frac{8}{3}\right)-\frac{1}{12a^{2}}+O\left(\frac{1}{a^{4}}\right)~.
   \end{split}
\end{align}
Similarly, from~\eqref{eq:FormualGq3n3NB1NC1Smalla} we find that the coefficient of the leading term depends directly on the subsystem size, with the corresponding $N\sim N_{A}\rightarrow \infty$ expectation for the behavior of $\textrm{GM}_{3}^{(\mathtt{3})}$ subject to the condition $N_{B}=N_{C}=1$, given by
\begin{align}
    \label{eq:FormualGq3n3NB1NC1SmallaLargeNA}
    \begin{split}
   \text{GM}^{(\mathtt{3})}_{3}(a,N_A)\Big\vert^{(a-1\ll 1~,N_{B}=N_{C}=1)}_{N_{A}\rightarrow \infty}\rightarrow -\frac{3}{16}(a-1)^{2}~.
   \end{split}
\end{align}
These results can be seen as representing the ``early-time'' behavior of the Rényi genuine multi-entropy curves discussed in Sec.~\ref{sec:GaussianMECurve}. In particular, ~\eqref{eq:FormualGq3n3NB1NC1SmallaLargeNA} would represent the early time behavior of the large $N_{A}$ limit of the top-left plot of Figure~\ref{fig:IndividualPlotsGMq3n3Page} corresponding to $a=11/10$. Similarly, ~\eqref{eq:FormualGq3n3NB1NC1LargeaLargeNA} would represent the early time behavior of the large $N_{A}$ limit of the bottom plots of Figure~\ref{fig:IndividualPlotsGMq3n3Page} corresponding to $a=10,100$. It would be very interesting to obtain similar analytic formulas for the subsystem configurations corresponding to the Page and Multi-entropy times in the large $N$ regime. This would be particularly important to determine the behavior of $\mathrm{GM}_{3}^{\mathtt{3}}$ in other relevant subsystem configurations.

On a similar vein, the results shown in Section~\ref{sec:GaussianMECurve} highlight the fact that in general, the Rényi genuine multi-entropy $\mathrm{GM}_{n}^{(\mathtt{3})}$ for $n>2$ cannot serve as a measure of entanglement for multi-party state and rather belongs to a class of quantities called signals~\cite{Balasubramanian:2024ysu,Gadde:2026msg}, as mentioned in Sec.~\ref{subsec:GHZstate}. This is because, in contrast with similar studies of the multi-entropy curve of Haar random states in holography~\cite{Iizuka:2024pzm,Iizuka:2025ioc}, the multi-entropy curves for $\mathrm{GM}_{3}^{(\mathtt{3})}(a)$ shown in Figs.~\ref{fig:PlotGMq3n3AllaPage} and~\ref{fig:IndividualPlotsGMq3n3Page} are negative for any value of $a$. In light of the asymptotic result~\eqref{AsymB} and $\mathrm{GM}_{2}^{(\mathtt{3})}=0$, we might wonder whether only the limit $n\rightarrow 1$ of the Rényi genuine multi-entropy could be positive.

Some possible future directions include the following:
\begin{itemize}
    \item Extend our results to \emph{fermionic} Gaussian states. Both bosonic and fermionic Gaussian states can be uniquely characterized by their Kähler structures~\cite{Hackl:2020ken}. It would be interesting to identify the differences between them and in particular verify whether $\textrm{GM}_{2}^{(3)}=0$ also holds for tripartite fermionic Gaussian states. We would also like to contrast this with the general CFT prediction $\Delta S^{(\mathtt{3})}_{2}=c\log(2)/4$ discussed in~\cite{Harper:2024ker,Harper:2025uui}.
    
    \item In a similar vein, it would be interesting to extend our multi-entropy curve results to ensembles of \emph{random} Gaussian states, more closely following~\cite{Bianchi:2021lnp}. This would require a use of the full machinery of Gaussian states as well as of random matrix theory.
    
    \item Similarly to multi-entropy, we can consider other multi-invariants such as the dihedral measure~\cite{Gadde:2024taa}, which has similarly interesting properties in free $(1+1)-$dimensional CFTs~\cite{Harper:2025uui}. 
    Note that, since $\Delta \mathcal{D}_{4}=\Delta S^{(3)}_{2}/2$, a similar vanishing of $\Delta \mathcal{D}_{4}=0$ occurs for ground state vacuum subregions of the free scalar theory~\cite{Harper:2025uui}.
    
    \item Authors in~\cite{Hu:2026bhg} highlight a peculiarity of the geometric properties of the $n=2$ multi-entropy $S^{(\mathtt{q})}_{n}$ in random tensor network (RTN) states, by showing that it is in general determined by minimal multiway cuts through the network, regardless of the number $\mathtt{q}$ of parties in the RTN state. They also show that for $n>2$ this particular minimal multiway cut property is not true in general. This observation that $n=2$ is special motivates the question of whether a similar special factorization of replica partition functions also holds for $\mathtt{q}$-partite general pure bosonic Gaussian states with $\mathtt{q}> 3$.

    \item 
    
    The factorization of $n=2$ replica partition functions is also observed in  stabilizer states \cite{Iizuka:2025pqq,Akella:2026xza} and in Abelian Chern–-Simons Theory \cite{Yuan:2025dgx,Chang:2026uds}. It is interesting to explore a unified understanding of why such factorization occurs in those states, including Gaussian states.
    
\end{itemize}



\section*{Acknowledgments}
\label{sec:Acknowledgments}
We are grateful to Jonathan Harper, Jaydeep Kumar Basak and Ali Mollabashi for valuable discussions, comments on the draft, and correspondence. We are also grateful to the organizers of the NCTS Taiwan String Workshop 2025 held at Hsinchu, National Tsing Hua University on December $1-5$, 2025, where this work was initiated. H.~A. Camargo was supported by the National Science and Technology Council, the Ministry of Education (Higher Education Sprout Project NTU-114L104022-1), and the National Center for Theoretical Sciences of Taiwan. M.~N.~was supported by JSPS KAKENHI Grant No.~JP25K23430.

\appendix


\section{Matrices for the replica partition functions}\label{app1}

Like the convention in \cite{Harper:2025uui},
let us decompose the $N_A+N_B$ dimensional matrices $\mathbf{K},\mathbf{L}$ in eq.~\eqref{eq:MatricesRedDMatAB} as follows
\begin{align}
    \mathbf{K}=\begin{pmatrix}
\mathbf{K}_{AA} & \mathbf{K}_{AB} \\
\mathbf{K}_{BA} & \mathbf{K}_{BB}
 \end{pmatrix},\;\;\;\mathbf{L}=\begin{pmatrix}
\mathbf{L}_{AA} & \mathbf{L}_{AB} \\
\mathbf{L}_{BA} & \mathbf{L}_{BB}
 \end{pmatrix}.
\end{align}
Here, $\mathbf{K}_{AA},\mathbf{L}_{AA}$ are $N_A\times N_A$ matrices that act on $A$, and $\mathbf{K}_{BB},\mathbf{L}_{BB}$ are $N_B\times N_B$ matrices that act on $B$. For the off-diagonal components, $\mathbf{K}_{AB},\mathbf{L}_{AB}$ are $N_A\times N_B$ matrices, and $\mathbf{K}_{BA},\mathbf{L}_{BA}$ are $N_B\times N_A$ matrices, where they couple $A$ to $B$. From these matrices, the $N_A+N_B$ dimensional matrices in eq.~\eqref{Matrices} are given by
\begin{align}
\begin{split}
     M_{A1}&=\frac{1}{2}\begin{pmatrix}    \mathbf{K}_{AA} + \mathbf{K}^*_{AA}& \mathbf{K}_{AB} +\dfrac{\mathbf{L}_{AB}+ \mathbf{L}^*_{AB}}{2}\\ \mathbf{K}_{BA} + \dfrac{\mathbf{L}_{BA}+ \mathbf{L}^*_{BA}}{2}\;\;\; &  \mathbf{K}_{BB} + \mathbf{K}^*_{BB}+\mathbf{L}_{BB} + \mathbf{L}^*_{BB}\end{pmatrix},\\
  M_{A2}&=\frac{1}{2}\begin{pmatrix}    \mathbf{L}_{AA}+ \mathbf{L}^*_{AA}& \mathbf{K}^*_{AB} +\dfrac{\mathbf{L}_{AB}+ \mathbf{L}^*_{AB}}{2}\\ \mathbf{K}^*_{BA} + \dfrac{\mathbf{L}_{BA}+ \mathbf{L}^*_{BA}}{2}\;\;\;& 0\end{pmatrix},
  \end{split}\\
  \begin{split}
   M_{B1}&=\frac{1}{2}\begin{pmatrix}     \mathbf{K}_{AA} + \mathbf{K}^*_{AA}+\mathbf{L}_{AA} + \mathbf{L}^*_{AA}\;\;\;& \mathbf{K}_{AB} +\dfrac{\mathbf{L}_{AB}+ \mathbf{L}^*_{AB}}{2}\\ \mathbf{K}_{BA} + \dfrac{\mathbf{L}_{BA}+ \mathbf{L}^*_{BA}}{2} &\mathbf{K}_{BB} + \mathbf{K}^*_{BB}\end{pmatrix},\\
   M_{B2}&=\frac{1}{2}\begin{pmatrix}    0& \mathbf{K}^*_{AB} +\dfrac{\mathbf{L}_{AB}+ \mathbf{L}^*_{AB}}{2}\\ \mathbf{K}^*_{BA} + \dfrac{\mathbf{L}_{BA}+ \mathbf{L}^*_{BA}}{2}\;\;\;& \mathbf{L}_{BB}+ \mathbf{L}^*_{BB}\end{pmatrix},
   \end{split}\\
   \begin{split}
   M_{C1}&=\frac{1}{2}\begin{pmatrix}     \mathbf{K}_{AA} + \mathbf{K}^*_{AA}\;\;\; &  \mathbf{K}_{AB} + \mathbf{K}^*_{AB} \\ \mathbf{K}_{BA} + \mathbf{K}^*_{BA}\;\;\; & \mathbf{K}_{BB} + \mathbf{K}^*_{BB} \end{pmatrix},\;\;\;
   M_{C2}=\frac{1}{2}\begin{pmatrix}    \mathbf{L}_{AA}+ \mathbf{L}^*_{AA}\;\;\;& \mathbf{L}_{AB}+ \mathbf{L}^*_{AB}\\  \mathbf{L}_{BA}+ \mathbf{L}^*_{BA}\;\;\; & \mathbf{L}_{BB}+ \mathbf{L}^*_{BB}\end{pmatrix},
   \end{split}\\
   \begin{split}\label{MM}
   M_{M1}&=\frac{1}{2}\begin{pmatrix}     \mathbf{K}_{AA} + \mathbf{K}^*_{AA}\;\;\; &  \mathbf{K}_{AB}   \\  \mathbf{K}_{BA}   & \mathbf{K}_{BB} + \mathbf{K}^*_{BB} \end{pmatrix},\;\;\;
M_{M2}=\frac{1}{2}\begin{pmatrix}    \mathbf{L}_{AA}+ \mathbf{L}^*_{AA}\;\;\;& \dfrac{\mathbf{L}_{AB}+ \mathbf{L}^*_{AB}}{2}\\  \dfrac{\mathbf{L}_{BA}+ \mathbf{L}^*_{BA}}{2} & 0\end{pmatrix},\\
M_{M3}&=\frac{1}{2}\begin{pmatrix}    0& \dfrac{\mathbf{L}_{AB}+ \mathbf{L}^*_{AB}}{2}\\  \dfrac{\mathbf{L}_{BA}+ \mathbf{L}^*_{BA}}{2}\;\;\; & \mathbf{L}_{BB}+ \mathbf{L}^*_{BB}\end{pmatrix},\;\;\;
M_{M4}=\frac{1}{2}\begin{pmatrix}     0 &  \mathbf{K}^*_{AB}  \\  \mathbf{K}^*_{BA}  & 0 \end{pmatrix}.
\end{split}
\end{align}
Note that, if $\mathbf{W}$ is real $\mathbf{W}^*=\mathbf{W}$, eq.~\eqref{MM} reduces to the formula in \cite{Harper:2025uui}. 


\section{Permutation-Symmetric Quadratic Hamiltonians}\label{app2}

Here we provide further details on permutation-symmetric quadratic Hamiltonians and show that they admit fully symmetric Gaussian states as ground states. We then discuss the permutation symmetry and finally we compare its properties with the Hamiltonian of a one-dimensional harmonic chain.

\subsection{Ground State}
\label{app2gs}

Consider the Hamiltonian~\eqref{eq:FreeHamSymm} for $N$ bosonic modes. We will show that it admits the state $\psi(\mathbf{q})=$ $(\det(\mathbf{W}/\pi))^{1/4}$ $\exp\left[-\frac{1}{2}\mathbf{q}^{\top}\cdot \mathbf{W}\cdot \mathbf{q}\right]$ as a ground state. Define the $N-$dimensional annihilation operator
\begin{align}
    \label{eq:AnnOp}
    \hat{\mathbf{b}}:=\frac{1}{\sqrt{2}}\left(\mathbf{W}^{1/2}\cdot\hat{\mathbf{q}}+i\mathbf{W}^{-1/2}\cdot\hat{\mathbf{p}}\right)~,
\end{align}
where $[\hat{b}_{j},\hat{b}^{\dagger}_{k}]=\delta_{jk}$. Acting with $\hat{\mathbf{p}}=-i\nabla_{\mathbf{q}}$ on $\psi(\mathbf{q})=\langle \mathbf{q}\vert \psi \rangle$ we have $i\hat{\mathbf{p}}\psi=\nabla_{\mathbf{q}} \psi =-(\mathbf{W}\cdot \hat{\mathbf{q}})\psi$. Thus
\begin{align}
    \label{eq:GroundStateEq}
    \hat{\mathbf{b}}\psi=\frac{1}{\sqrt{2}}\left(\mathbf{W}^{1/2}\cdot\hat{\mathbf{q}}+\mathbf{W}^{-1/2}\cdot(i\hat{\mathbf{p}})\right)\psi=\frac{1}{\sqrt{2}}\left(\mathbf{W}^{1/2}\cdot\hat{\mathbf{q}}-\mathbf{W}^{-1/2}\cdot\mathbf{W}\cdot \hat{\mathbf{q}}\right)\psi=0~.
\end{align}
Moreover, we can re-write the Hamiltonian~\eqref{eq:FreeHamSymm} in terms of creation and annihilation operators as 
\begin{align}
    \label{eq:FreeHamSymAnnOp}
    \mathbf{H}=\frac{1}{2}\left(\hat{\mathbf{p}}^{\top}\cdot\mathbf{W}^{-1}\cdot\hat{\mathbf{p}}+\hat{\mathbf{q}}^{\top}\cdot \mathbf{W}\cdot\hat{\mathbf{q}}\right)\mapsto H=\sum_{j=1}^{N}\hat{b}_{j}^{\dagger}\hat{b}_{j}+\frac{N}{2}~.
\end{align}
This Hamiltonian can also be written in the canonical form
\begin{align}
    \label{eq:FreeHamSymCanon}
    \mathbf{H}=\frac{1}{2}\left(\hat{\mathbf{P}}^{\top}\cdot\hat{\mathbf{P}}+\hat{\mathbf{Q}}^{\top}\cdot\hat{\mathbf{Q}}\right)
\end{align}
under the canonical change of variables
\begin{align}
    \label{eq:CanVariables}
    \hat{\mathbf{P}}=\mathbf{W}^{-1/2}\cdot \hat{\mathbf{p}}~,\quad \hat{\mathbf{Q}}=\mathbf{W}^{1/2}\cdot \hat{\mathbf{q}}~.
\end{align}

\subsection{Permutation Symmetry}
\label{app2permut}
A permutation $\sigma \in S_{N}:=\lbrace 1,\ldots,N\rbrace$ can be represented by a $N\times N$ permutation matrix $\mathbf{P}_{\sigma}$ whose components are
\begin{align}
    \label{eq:PermMat}
(\mathbf{P}_{\sigma})_{ij}:=\delta_{i,\sigma(j)}~.
\end{align}
Acting on a $N-$dimensional vector $\mathbf{x}=(x_{1},\ldots,x_{N})^{\top}$ simply yields
\begin{align}
    \label{eq:ActionPermMat}
    x'_{i}=(\mathbf{P}_{\sigma}\cdot \mathbf{x})_{i}=x_{\sigma(i)}~.
\end{align}
That is, $\mathbf{P}_{\sigma}$ reorders the components of vectors according to the permutation $\sigma$. In general, permutations are \emph{orthogonal} $\mathbf{P}_{\sigma}^{\top}\cdot \mathbf{P}_{\sigma}=\mathbb{I}_{N}$, have the \emph{group representation} $\mathbf{P}_{\sigma_{1}}\cdot \mathbf{P}_{\sigma_{2}}=\mathbf{P}_{(\sigma_{1}\circ\sigma_{2})}$ and define a \emph{conjugation} of matrices $\mathbf{M}\mapsto\mathbf{P}_{\sigma}^{\top}\cdot\mathbf{M}\cdot \mathbf{P}_{\sigma}$ which permutes the rows and columns of $\mathbf{M}$ consistently with $\sigma$. In particular, any matrix $\mathbf{M}$ built from the matrices $\mathbb{I}_{N}$ and $\mathbf{J}_{N}$ where $\mathbb{I}_{N}$ is the $N\times N$ identity matrix and $\mathbf{J}_{N}$ is the $N\times N$ matrix with entries $(\mathbf{J}_{N})_{ij}=1\quad\forall\,i,j=1,\ldots,N$ according to
\begin{align}
    \label{eq:ActionPmatMat}
    \mathbf{M}=\alpha \mathbb{I}_{N}+\beta \mathbf{J}_{N}~,
\end{align}
is invariant under the action of $\mathbf{P}_{\sigma}$, since
\begin{align}
    \label{eq:PermutIandJ}
   \mathbf{P}_{\sigma}^{\top}\cdot \mathbf{P}_{\sigma}=\mathbb{I}_{N}~, \quad~\,\mathbf{P}_{\sigma}^{\top}\cdot\mathbf{J}_{N}\cdot \mathbf{P}_{\sigma}=\mathbf{J}_{N}~.
\end{align}
This means that any such $\mathbf{M}$ commutes with every permutation matrix $[\mathbf{M},\mathbf{P}_{\sigma}]=0~,\quad\forall \sigma \in S_{N}$ for any $\alpha, \beta$.

Under the action of a permutation matrix $\mathbf{P}_{\sigma}$, the canonical coordinates $\mathbf{q}=(q^{1},\ldots,q^{N})^{\top}$ and momenta $\mathbf{p}=(p_{1},\ldots,p_{N})^{\top}$ transform according to $\mathbf{q}\mapsto\mathbf{q}'=\mathbf{P}_{\sigma}\cdot \mathbf{q}$ (i.e. $q^{'i}=q^{\sigma(i)}$) and $\mathbf{p}\mapsto\mathbf{p}'=\mathbf{P}_{\sigma}\cdot \mathbf{p}$ (i.e. $p'_{i}=p_{\sigma(i)}$) so that the Hamiltonian~\eqref{eq:FreeHamSymAnnOp} is invariant under this change of canonical coordinates
\begin{align}
    \mathbf{H}\mapsto\mathbf{H}'=\frac{1}{2}\left(\hat{\mathbf{p}}^{\top}\cdot\mathbf{P}_{\sigma}^{\top}\cdot\mathbf{W}^{-1}\cdot\mathbf{P}_{\sigma}\cdot\hat{\mathbf{p}}+\hat{\mathbf{q}}^{\top}\cdot \mathbf{P}_{\sigma}^{\top}\cdot\mathbf{W}\cdot\mathbf{P}_{\sigma}\cdot\hat{\mathbf{q}}\right)=\mathbf{H}~,
\end{align}
since $\mathbf{P}_{\sigma}^{\top}\cdot\mathbf{W}\cdot\mathbf{P}_{\sigma}=\mathbf{W}$ and $\mathbf{P}_{\sigma}^{\top}\cdot\mathbf{W}^{-1}\cdot\mathbf{P}_{\sigma}=\mathbf{W}^{-1}$. This means that this Hamiltonian is invariant under the full symmetric group $S_{N}$.

\subsection{Comparison with Harmonic Chain}
\label{app2HarmChain}

We can also compare directly this Hamiltonian with the standard harmonic chain Hamiltonian. The discretized version of the Hamiltonian of a massive $(1+1)-$dimensional scalar field on $N$ sites with periodic boundary conditions $q_{N+1}=q_{1}$, mass $m$ and lattice spacing $\delta$ is given by
\begin{align}
    \label{eq:HamChain}
    H_{\textrm{chain}}=\frac{\delta}{2}\sum_{j=0}^{N-1}\left(p_{j}^{2}+\frac{m^{2}}{\delta^{2}}q_{j}^{2}+\frac{1}{\delta^{4}}(q_{j}-q_{j+1})^{2}\right)\mapsto\frac{\delta}{2}\left(\hat{\mathbf{p}}^{\top}\cdot\hat{\mathbf{p}}+\hat{\mathbf{q}}^{\top}\cdot \mathbf{V}_{\textrm{circ}}\cdot\hat{\mathbf{q}}\right)\equiv \mathbf{H}_{\textrm{chain}}~,
\end{align}
where the potential $\mathbf{V}_{\textrm{circ}}$ is a circulant matrix with diagonal entries $m^{2}+2/\delta^{2}$ and nearest-neighbor off-diagonals $-1/\delta^{2}$, $\mathbf{V}_{\textrm{circ}}=(\frac{1}{\delta^{2}})\mathrm{circ}\left((m\delta)^{2}+2,-1,0,\ldots,0,-1\right)$ . This Hamiltonian can be diagonalized by introducing discrete Fourier modes
\begin{align}
    \label{eq:FourierHarmChain}
    q_{j}:=\frac{1}{\sqrt{N}}\sum_{k=0}^{N-1}e^{\frac{2\pi ik}{N}j}\phi_{k}~,\quad p_{j}:=\frac{1}{\sqrt{N}}\sum_{k=0}^{N-1}e^{-\frac{2\pi ik}{N}j}\pi_{k}~.
\end{align}
leading to the Hamiltonian 
\begin{align}
    \label{eq:HamChainFourier}
    H_{\textrm{chain}}=\frac{\delta}{2}\sum_{k=0}^{N-1}\left(\vert \pi_{k}\vert^{2}+\omega_{k}^{2}\vert \phi_{k}\vert^{2}\right)~,
\end{align}
where
\begin{align}
    \label{eq:FreqsHarmChain}
    \omega_{k}^{2}=m^{2}+\frac{4}{\delta^{2}} \sin^{2}\left(\frac{\pi k}{N}\right)~, \quad k=0,\ldots,N-1~.
\end{align}
The correlators for the ground state $\vert 0 \rangle$ of~\eqref{eq:HamChain} in Fourier space are given by
\begin{align}
\label{eq:HarmChainFourierCorr}
    \langle \tilde{\phi}_{k}\tilde{\phi}_{-k}\rangle=\frac{1}{2\omega_{k}}~,\quad  \langle \tilde{\pi}_{k}\tilde{\pi}_{-k}\rangle=\frac{\omega_{k}}{2}~,\quad \langle \tilde{\phi}_{k}\tilde{\pi}_{k'}+\tilde{\pi}_{k'}\tilde{\phi}_{k}\rangle=0~,
\end{align}
with $\tilde{\phi}_{k}=\sqrt{\delta}\,\phi_{k}$ and $\tilde{\pi}_{k}=\sqrt{\delta}\,\pi_{k}$, so that $[\tilde{\phi}_{k},\tilde{\pi}_{k'}]=i\delta_{k,k'}$, and in position space they become
\begin{align}
\label{eq:HarmChainPositionCorr}
    \langle \tilde{q}_{i}\tilde{q}_{j}\rangle=\frac{1}{2N}\sum_{k=0}^{N-1}\frac{e^{\frac{2\pi i k}{N}(i-j)}}{\omega_{k}}~,\quad  \langle \tilde{p}_{i}\tilde{p}_{j}\rangle=\frac{1}{2N}\sum_{k=0}^{N-1}\omega_{k}e^{\frac{2\pi i k}{N}(i-j)}~,\quad \langle \tilde{q}_{i}\tilde{p}_{j}+\tilde{p}_{j}\tilde{q}_{i}\rangle=0~,
\end{align}
where $\tilde{q}_{i}=\sqrt{\delta}\,q_{i}$ and $\tilde{p}_{i}=\sqrt{\delta}\,p_{i}$. This gives the single-site (on-site) variances
\begin{align}
\label{eq:HarmChainPositionVariance}
    \langle \tilde{q}_{i}^{2}\rangle=\frac{1}{2N}\sum_{k=0}^{N-1}\frac{1}{\omega_{k}}~,\quad  \langle \tilde{p}_{i}^{2}\rangle=\frac{1}{2N}\sum_{k=0}^{N-1}\omega_{k}~,
\end{align}
and the reduced single-site covariance matrix
\begin{align}
    \label{eq:SingleSiteCovarianceMatrix}
    \sigma_{\mathrm{site}}=\begin{pmatrix}
        \langle \tilde{q}_{i}^{2}\rangle& 0 \\
        0& \langle \tilde{p}_{i}^{2}\rangle\\
    \end{pmatrix}
\end{align}
corresponding to the single-site density operator $\rho_{i}$ whose single-site purity $\mu$ is given by
\begin{align}
    \label{eq:HarmChainSitePurity}
    \mu_{\mathrm{site}}:=\mathrm{tr}(\rho_{i}^{2})=\frac{1}{2\sqrt{\det(\sigma_{\mathrm{site}})}}=\frac{1}{2\sqrt{\langle \tilde{q}_{i}^{2}\rangle\langle \tilde{p}_{i}^{2}\rangle}}=\frac{N}{\sqrt{\left(\sum_{k=0}^{N-1}\frac{1}{\omega_{k}}\right)\left(\sum_{k=0}^{N-1} \omega_{k}\right)}}~.
\end{align}
In the thermodynamic limit, taking $N\rightarrow \infty$ while keeping the circumference $L=N\delta$ fixed, this becomes
\begin{align}
    \label{eq:HarmChainSitePurityThermo}
    \mu_{\mathrm{site}}^{(\mathrm{th})}=\frac{N}{\sqrt{\langle \omega\rangle_{\mathrm{BZ}} \langle \omega^{-1}\rangle}_{\mathrm{BZ}}}~,\quad \langle f(\omega)\rangle_{\mathrm{BZ}}:=\frac{\delta}{2\pi}\int_{-\pi/\delta}^{\pi/\delta}\mathrm{d}k\,f(\omega(k))~,
\end{align}
with $\omega(k)=\sqrt{m^{2}+(4/\delta^{2})\sin^{2}(k\delta /2)}$, where we converted the sum in the lattice with spacing $\delta$ into the integral in first Brillouin zone (BZ) by identifying $k\mapsto k+2\pi/\delta$ with $k\in [-\pi/\delta , \pi/\delta]$. From this expression we see that in the large mass limit $m\delta \gg 1$, the individual frequencies become approximately independent of $k$, $\omega_{k}\approx m$, $\forall k$. In this case, we have $\sum_{k} \omega_{k}\approx Nm$ and $\sum_{k} \omega_{k}^{-1}\approx N/m$ so that $\mu_{\mathrm{site}}^{(\mathrm{th})} \rightarrow 1$ in this limit, which means that for a scalar field with large mass, the single-site mixed states become nearly pure and decoupled from each other. In contrast, in the small mass limit $m\delta \ll 1$ we have $\omega(k)\sim (2/\delta)\vert\sin(k\delta/2)\vert$, which vanishes linearly near $k=0$. This gives a logarithmic divergence for $\int \mathrm{d}k \,(1/\omega(k))$ and consequently, $\mu_{\mathrm{site}}^{(\mathrm{th})} \rightarrow 0$ in this limit. This means that small frequency (long wavelength) fluctuations make the single-site state $\rho_{i}$ highly mixed. This intuitively gives the following dependence of the purity $\mu_{\mathrm{site}}^{(\mathrm{th})}$ of the single-site on the dimensionless combination $m\delta$
\begin{align}
\label{eq:HarmChainPurityMass}
    \mu_{\mathrm{site}}^{(\mathrm{th})}\rightarrow \begin{cases}
        0~\,\quad \mathrm{for}\,\,m\delta\ll 1~,\\
        1~\,\quad \mathrm{for}\,\,m\delta\gg 1~,
    \end{cases}~.
\end{align}
In contrast, the permutation symmetric Hamiltonian~\eqref{eq:FreeHamSymm} can be written in components as
\begin{align}
    \label{eq:FreeHamSymmComponents}
    \mathbf{H}=\frac{1}{2}\left(\frac{1}{\alpha}\sum_{i=1}^{N}p_{i}^{2}+\alpha\sum_{i=1}^{N}q_{i}^{2}-\frac{e^{-}}{\alpha(\alpha+Ne^{-})}\left(\sum_{i=1}^{N}p_{i}\right)^{2}+e^{-}\left(\sum_{i=1}^{N}q_{i}\right)^{2}\right)~.
\end{align}
The last two terms encode the \emph{non-locality} of this Hamiltonian with all-to-all potential and kinetic interactions. We can explicitly perform a normal-mode decomposition by projecting onto the symmetric center-of-mass subspace. Define the projection operator
\begin{align}
    \label{eq:Proj}
    \mathbf{\Pi}:=\frac{\mathbf{J}_{N}}{N}~,\quad \mathbf{\Pi}^{2}=\mathbf{\Pi}~,\quad\mathbf{\Pi}^{\top}=\mathbf{\Pi}~.
\end{align}
Then,
\begin{align}
    \label{eq:ProjWmat}
    \mathbf{W}=\alpha \mathbb{I}_{N}+e^{-}\mathbf{J}_{N}=\alpha(\mathbb{I}_{N}-\mathbf{\Pi})+(\alpha+Ne^{-})\mathbf{\Pi}~.
\end{align}
This shows that $(\mathbb{I}_{N}-\mathbf{\Pi})$ and $\mathbf{\Pi}$ are orthogonal projectors onto complementary subspaces. The inverse of $\mathbf{W}$ is given by
\begin{align}
    \label{eq:WinvProj}
    \mathbf{W}^{-1}=\frac{1}{\alpha}(\mathbb{I}_{N}-\mathbf{\Pi})+\frac{1}{\alpha+Ne^{-}}\mathbf{\Pi}~.
\end{align}
Eqs.~\eqref{eq:ProjWmat} and~\eqref{eq:WinvProj} make the spectral decomposition manifest, in which there is a $1-$dimensional symmetric subspace corresponding to the eigenvalue $\lambda_{\mathrm{sym}}=\alpha+Ne^{-}$, and a $(N-1)-$dimensional transverse subspace corresponding to the eigenvalue $\lambda_{\perp}=\alpha$. Define the center-of-mass position and momenta
\begin{align}
    Q_{s}=\frac{1}{\sqrt{N}}\sum_{i=1}^{N}q_{i}~,\quad P_{s}=\frac{1}{\sqrt{N}}\sum_{i=1}^{N}p_{i}~,
\end{align}
and define $N-1$ orthonormal transverse coordinates $\lbrace Q_{\perp,j},P_{\perp,j}\rbrace$ with $\sum_{j=1}^{N-1}Q_{\perp,j}=0=\sum_{j=1}^{N-1}P_{\perp,j}$. In these coordinates, the Hamiltonian reads
\begin{align}
    \label{eq:FreeHamSymmCoM}
    \mathbf{H}=\frac{1}{2}\left(\kappa_{\mathrm{sym}}P_{s}^{2}+\lambda_{\mathrm{sym}}Q_{s}^{2}+\sum_{j=1}^{N-1}\left(\kappa_{\perp}P_{\perp,j}^{2}+\lambda_{\perp}Q_{\perp,j}^{2}\right)\right)~,
\end{align}
where $\kappa_{\perp}=1/\alpha$ and $\kappa_{\mathrm{sym}}=1/(\alpha+Ne^{-})$. Thus, this model separates into one collective oscillator in the center of mass and $N-1$ degenerate orthogonal oscillators.

\bibliographystyle{JHEP}
\bibliography{references}  

\end{document}